\documentclass[final,5p,times,twocolumn,number]{elsarticle}

\usepackage[T1]{fontenc}
\usepackage[utf8]{inputenc}

\makeatletter
\def\ps@pprintTitle{%
  \let\@oddhead\@empty
  \let\@evenhead\@empty
  \let\@oddfoot\@empty
  \let\@evenfoot\@oddfoot
}
\makeatother

\usepackage{booktabs}
\usepackage{array}
\usepackage{tabularx}
\usepackage{amssymb}
\usepackage{amsmath}
\usepackage{flushend}
\usepackage{placeins}
\usepackage{stfloats}
\usepackage{xcolor}
\usepackage{hyperref}
\usepackage{graphicx}
\usepackage{newunicodechar}
\usepackage{enumitem}

\hypersetup{colorlinks=true,linkcolor=blue,urlcolor=blue,citecolor=blue}
\setlist[itemize]{itemsep=1pt,topsep=2pt,leftmargin=*}
\newunicodechar{²}{\ensuremath{^2}}
\newunicodechar{·}{\ensuremath{\cdot}}
\newunicodechar{×}{\ensuremath{\times}}
\newunicodechar{−}{\ensuremath{-}}
\newunicodechar{Δ}{\ensuremath{\Delta}}
\newunicodechar{γ}{\ensuremath{\gamma}}
\newunicodechar{δ}{\ensuremath{\delta}}
\newunicodechar{θ}{\ensuremath{\theta}}
\newunicodechar{λ}{\ensuremath{\lambda}}
\newunicodechar{ν}{\ensuremath{\nu}}
\newunicodechar{ρ}{\ensuremath{\rho}}
\newunicodechar{σ}{\ensuremath{\sigma}}
\newunicodechar{ᵀ}{\ensuremath{^{\top}}}
\newunicodechar{₁}{\ensuremath{_1}}
\newunicodechar{₂}{\ensuremath{_2}}
\newunicodechar{→}{\ensuremath{\rightarrow}}
\newunicodechar{√}{\ensuremath{\surd}}
\newunicodechar{≈}{\ensuremath{\approx}}
\newunicodechar{≤}{\ensuremath{\leq}}
\newunicodechar{≥}{\ensuremath{\geq}}
\newunicodechar{≠}{\ensuremath{\neq}}
\newunicodechar{α}{\ensuremath{\alpha}}
\newunicodechar{Σ}{\ensuremath{\Sigma}}
\newunicodechar{κ}{\ensuremath{\kappa}}
\newunicodechar{τ}{\ensuremath{\tau}}
\newunicodechar{φ}{\ensuremath{\phi}}
\newunicodechar{ψ}{\ensuremath{\psi}}
\newunicodechar{ω}{\ensuremath{\omega}}
\newunicodechar{ᵗ}{\ensuremath{^{t}}}
\newunicodechar{⁴}{\ensuremath{^{4}}}
\newunicodechar{←}{\ensuremath{\leftarrow}}
\newunicodechar{⌈}{\ensuremath{\lceil}}
\newunicodechar{⌉}{\ensuremath{\rceil}}
\newunicodechar{⟨}{\ensuremath{\langle}}
\newunicodechar{⟩}{\ensuremath{\rangle}}
\newunicodechar{μ}{\ensuremath{\mu}}
\newunicodechar{≪}{\ensuremath{\ll}}
\newunicodechar{∞}{\ensuremath{\infty}}
\newunicodechar{→}{\ensuremath{\rightarrow}}
\usepackage[most]{tcolorbox}
\newtcolorbox{practinner}[1]{enhanced,sharp corners,colback=white,colframe=black,boxrule=0.5pt,
  colbacktitle=white,coltitle=black,titlerule=0pt,fonttitle=\bfseries\small,title={#1},
  attach title to upper={\ \ },left=5pt,right=5pt,top=4pt,bottom=4pt,fontupper=\small\setlength{\parskip}{4pt}}
\begin{document}
\begin{frontmatter}
\title{Network Effects in Randomized Experiments: Simulations at Scale}
\author{M\'arton Trencs\'eni (\texttt{mtrencseni@gmail.com})}

\begin{abstract}

On a social network, users post more when they see their friends post more. In an A/B test, this means that treating some users changes the behavior of control users as well, and the test no longer measures the effect of launching the treatment to everyone. This paper studies these effects on five families of random graphs, on graphs with up to $10^9$  nodes, for two response functions to content seen. We model each user's content production as $c = \lambda + \Delta\lambda z + \nu g(Ac)$: a baseline, plus the treatment effect, plus a response to what the user's friends post, either linear, $g(s) = s$, or logarithmic, $g(s) = K\log(1 + s/K)$, where each additional post seen adds less than the previous one. We compare the difference in means (DM) reported by the A/B test with the exact global average treatment effect (GATE) of a launch, and check a closed form for each effect. In the linear model a finite steady state exists only when $\nu\rho(A) < 1$, where $\rho(A)$  is the largest eigenvalue of the adjacency matrix. On heavy-tailed graphs a peer effect that looks safe by the mean degree becomes unstable as the graph grows. On graphs with homogeneous degrees the test reports about $1 - \nu\rho$  of the launch effect, running it longer does not close the gap, and cluster randomization narrows it. With the logarithmic response a finite steady state exists for every peer effect, and the test reports about $1 - \rho(J)$  of the launch effect, where $\rho(J) < 1$  is the largest eigenvalue of the matrix $J$  which describes how a small treatment spreads, and takes the place of $\nu\rho(A)$. Treatment that adds edges or users can make the test miss most of the effect.

\end{abstract}
\end{frontmatter}

\section{Introduction}

In a randomized experiment on a social network, the treatment of one user affects the outcomes of others. A treated user who posts more is seen by friends in both arms, and those friends post more in turn. The A/B test compares two groups that influence each other, so the difference between them is not the effect of launching the treatment to everyone. This is interference (Hudgens and Halloran~\cite{r13}). For linear outcome models, Eckles, Karrer and Ugander~\cite{r8} give the bias of the difference in means in closed form, restated in Section 2 for finite $N$. They also show that graph cluster randomization reduces the bias.\footnote{In graph cluster randomization the users are first partitioned into clusters of connected users, and each cluster is then assigned to treatment or control as a whole. A treated user's friends are then more often treated too, so fewer friendships cross between the two arms (Ugander, Karrer, Backstrom and Kleinberg~\cite{r21}).}

The existing literature gives the bias as a formula, but not its size on graphs that resemble social networks. Three features of social networks matter here: a user's friends are often friends of each other (clustering), users form groups with many friendships inside and few between (communities), and a few users have very many friends while most have few (a heavy-tailed degree distribution). This paper measures the gap between what the A/B test reports and the effect of launching the treatment to everyone on five random graph families: Watts-Strogatz graphs, which are clustered; stochastic block models (SBM), which have communities; Barabási-Albert and power-law graphs, which are heavy-tailed; and Erdős-Rényi graphs, which have none of the three (Table~\ref{tab:families}). It also measures how the gap changes with the size of the graph and over time after launch, and how much two remedies reduce it: treating whole groups of connected users together (cluster randomization), and estimators that compare outcomes at different treated fractions. The main model is linear: each user's production is linear in the production of the users they see. Section 8 repeats the main results with a logarithmic response, in which each additional post seen adds less than the previous one. In both models the true effect of a full launch can be computed numerically: on graphs with up to $10^9$  nodes in the linear model and $10^5$  nodes in the logarithmic one.

\begin{table*}[!tp]
\centering
\scriptsize
\renewcommand{\arraystretch}{1.15}
\begin{tabular*}{\linewidth}{@{}l@{\extracolsep{\fill}}ccccc@{}}
\toprule
\textbf{} & \textbf{Watts-Strogatz} & \textbf{Erdős-Rényi} & \textbf{SBM} & \textbf{Barabási-Albert} & \textbf{Power-law} \\
\midrule[\heavyrulewidth]
 & \raisebox{-0.5\height}{\includegraphics[height=1.7cm]{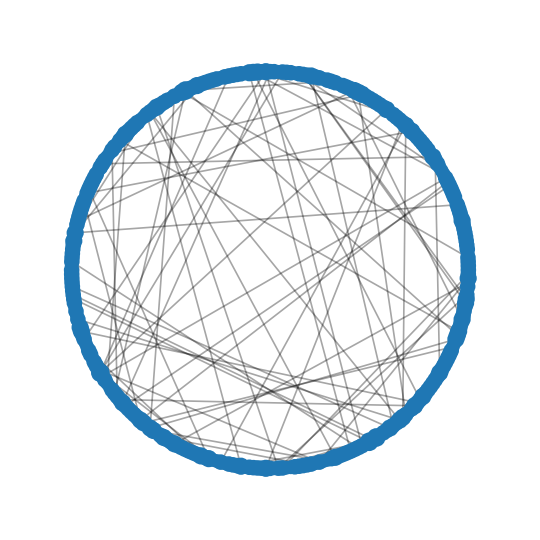}} & \raisebox{-0.5\height}{\includegraphics[height=1.7cm]{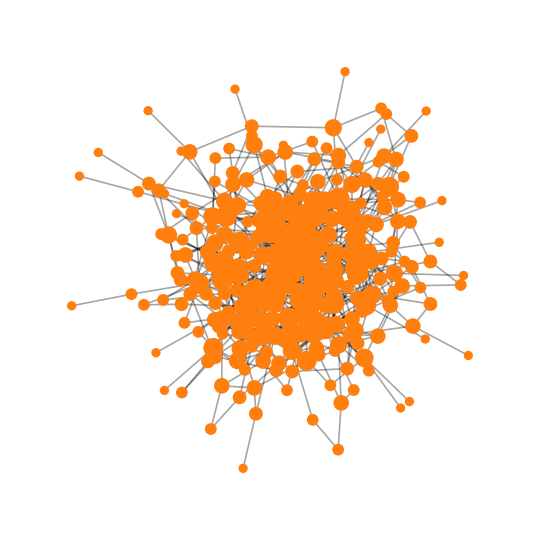}} & \raisebox{-0.5\height}{\includegraphics[height=1.7cm]{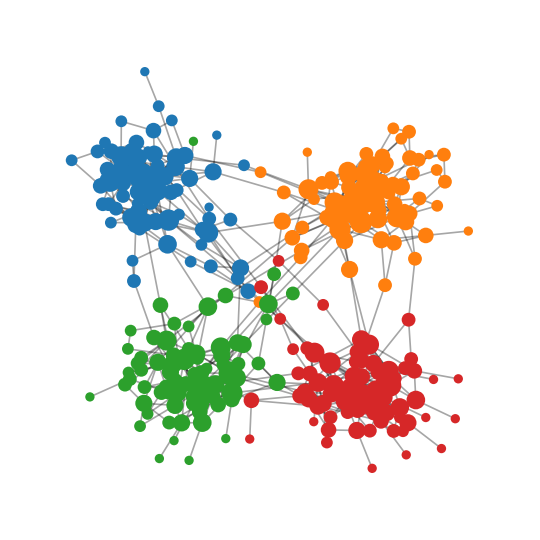}} & \raisebox{-0.5\height}{\includegraphics[height=1.7cm]{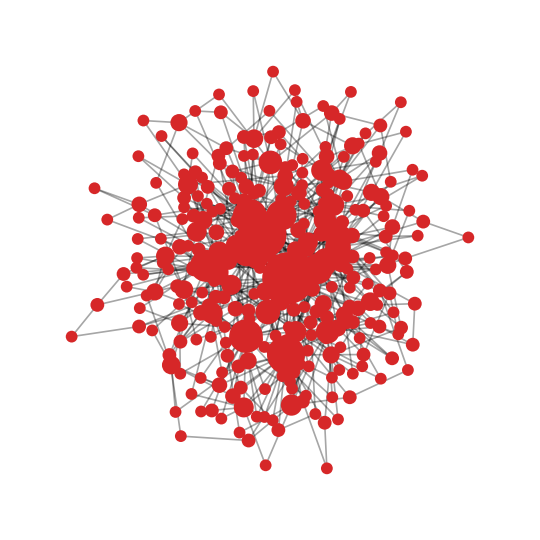}} & \raisebox{-0.5\height}{\includegraphics[height=1.7cm]{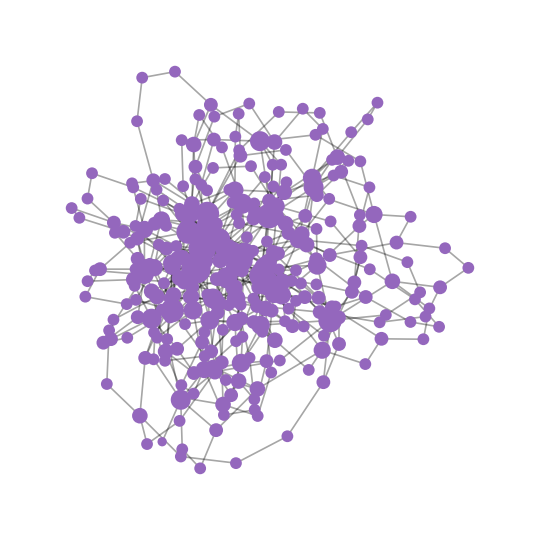}}\\
\hline
construction & \begin{tabular}[c]{@{}c@{}}ring lattice, \\ 10\% of edges rewired\end{tabular} & \begin{tabular}[c]{@{}c@{}}edges uniformly \\ at random\end{tabular} & \begin{tabular}[c]{@{}c@{}}100 blocks, 10\% of \\ edges between blocks\end{tabular} & \begin{tabular}[c]{@{}c@{}}preferential \\ attachment\end{tabular} & \begin{tabular}[c]{@{}c@{}}configuration model, \\ $\gamma = 2.5$, $k_{\max} = \sqrt N$ \end{tabular}\\
\hline
clustered & yes (0.52) & no (0.000) & no (0.015) & no (0.001) & no (0.001)\\
\hline
communities & no (0.83) & no (0.18) & yes (0.90) & no (0.17) & no (0.17)\\
\hline
heavy-tailed & no (0.07) & no (0.22) & no (0.22) & yes (1.41) & yes (1.23)\\
\hline
$\rho(A)/\bar k$  & 1.01 & 1.05 & 1.05 & 3.15 & 2.60\\
\hline
group & homogeneous & homogeneous & homogeneous & \textbf{heavy-tailed} & \textbf{heavy-tailed} \\
\bottomrule
\end{tabular*}
\vspace{4pt}
\caption{The five graph families, with $N = 10^5$  users and $\bar k = 20$  friends on average. The drawings are small examples with 300 users; node size grows with the number of friends. The numbers in parentheses are the share of a user's friends who are friends of each other (clustered), the share of friendships inside clusters of 1,000 users (communities), and the standard deviation of the number of friends over its mean (heavy-tailed). $\rho(A)$  is the largest eigenvalue of the adjacency matrix, which sets the stability limit $\nu\rho < 1$. Watts-Strogatz graphs have no communities by construction, but contiguous stretches of the ring keep most friendships inside clusters (Section 6.1).}\label{tab:families}
\end{table*}

The contributions are:

\begin{itemize}
\item the size of the bias on five graph families and on graphs with up to $10^9$  nodes;
\item how the launch effect and the A/B difference evolve after launch;
\item cluster designs and estimators compared against the exact launch effect, including near the stability limit;
\item treatment that changes the graph;
\item a concave response, under which a finite steady state exists for every peer effect and the test reports about $1 - \rho(J)$  of the launch effect, where $\rho(J) < 1$  describes how a small treatment spreads;
\item closed forms or approximations for every effect, with the conditions under which they fail.
\end{itemize}

An earlier paper~\cite{r20} simulated this model on Watts-Strogatz graphs and named several of its effects; Appendix D maps those names to the standard ones.

\section{Model and estimands}

Each user $i$  produces $c_i$  units of content per period.\footnote{For the mathematics of this section, see the companion tutorial \emph{A Mathematics Tutorial on Social Network Experimentation:} $\rho(A)$, \emph{GATE, ADE and E[DM] for Linear and Log Models}.} Production follows
\begin{equation}c^t = \lambda + \Delta\lambda\,z + \nu A c^{t-1}\label{eq:model}\end{equation}
each period a user produces a baseline $\lambda > 0$, plus $\Delta\lambda$  if treated ($z_i = 1$), plus $\nu \geq 0$  times the content they saw in the previous period. Row $i$  of the adjacency matrix $A$  lists the users whose content user $i$  sees. One period is one hop of influence.

If the iteration settles, we set $c = c^t = c^{t-1}$  and solve the equation for $c$. The steady-state solution is $c = M(\lambda + \Delta\lambda z)$  with $M = (I - \nu A)^{-1} = I + \nu A + (\nu A)^2 + \dots$. Each term is one more hop: $(\nu A)^t$  carries influence along paths of length $t$. The series converges if and only if $\nu\rho(A) < 1$, where $\rho(A)$  is the largest eigenvalue of $A$; otherwise production grows without bound, which is unrealistic. After many hops, the extra content is spread over users in proportion to the eigenvector $v$  of $\rho(A)$  with $v^{\top}v = 1$, the Perron vector: for large $t$, $(\nu A)^t \approx (\nu\rho)^t vv^{\top}$. On a graph where every user has the same number of friends, $v$  is flat and every user receives the same share; on a graph with hubs, $v$  is concentrated on the hubs, and so is the extra content. For undirected graphs $\rho(A) \geq \bar k$, with equality only for regular graphs, so $\nu\bar k < 1$  is necessary for stability but not sufficient.

Sections 4 and 5 use exact values, without observation noise. In Section 6, where the precision of an estimate is studied, the test is read over one week, after production has settled: for each user, $W = 7$  daily counts are drawn from a Poisson distribution with mean $\mu$  equal to the user's steady-state production per day, and summed.

The global average treatment effect (GATE) is the launch effect: the average production with everyone treated minus the average production with no one treated. Since $c(\mathbf{1}) - c(\mathbf{0}) = \Delta\lambda M\mathbf{1}$,
\begin{equation}\mathrm{GATE} = \Delta\lambda\,\mathbf{1}^{\top}M\mathbf{1}/N\label{eq:gate}\end{equation}
The average direct effect (ADE) is the change in a user's own production when only that user is treated, averaged over users. Treating only user $i$  changes $c_i$  by $\Delta\lambda M_{ii}$, so
\begin{equation}\mathrm{ADE} = \Delta\lambda\operatorname{tr}(M)/N\label{eq:ade}\end{equation}
It includes the content that returns to the treated user through their friends, which is small: $\mathrm{ADE} \approx \Delta\lambda$  when $\nu^2\bar k$  is small. We write $m = \mathbf{1}^{\top}M\mathbf{1}/N$  for the social multiplier, the GATE divided by $\Delta\lambda$, and $\tau = \operatorname{tr}(M)/N$  for the ADE divided by $\Delta\lambda$.

The A/B test reports the difference in means (DM): the mean production of the treated arm minus the mean production of the control arm. A single DM also contains the difference between the two arms that was there before treatment, which averages to zero over random assignments. Under Bernoulli assignment with both arms nonempty, the expected DM is exactly (derived in Section 4.2 up to terms of order $1/N$)
\begin{equation}E[\mathrm{DM}] = (N\,\mathrm{ADE} - \mathrm{GATE})/(N - 1)\label{eq:edm}\end{equation}
whatever the treated fraction. For large $N$  this is the ADE (Eckles, Karrer and Ugander~\cite{r8}, eq. 8); the correction matters only when the GATE is comparable to $N$  times the ADE, and on the graphs of this paper it is negligible. The ratio $E[\mathrm{DM}]/\mathrm{GATE}$  is the share of the launch effect the test reports, and the central quantity of this paper; we call it the attenuation ratio.

\section{Simulation setup}

Every simulation runs on the five graph families of Table~\ref{tab:families}. Unless stated otherwise, graphs have $N = 10^5$  nodes and mean degree $\bar k = 20$, the baseline production is $\lambda = 1$  and the treatment adds $\Delta\lambda = 0.05$, so the direct effect is 5\% of baseline production (Appendix C and Table~\ref{tab:setup}). The power-law configuration model has degree exponent $\gamma = 2.5$  and maximum degree $\sqrt N$. The peer effect is set by its distance from the stability limit, $\nu\rho$  from 0.1 to 0.999, except where a fixed $\nu\bar k$  is stated.

The ground truth comes from solving $(I - \nu A)c = b$  with a sparse solver to a relative tolerance of $10^{-10}$, for many assignments at once; "exact" below means this, as opposed to a closed-form approximation. Each point averages several graphs and treatment assignments (Table~\ref{tab:setup}), and error bars are standard errors over graphs. The simulator (the \texttt{netfx} package, Appendix C), the raw results and the scripts that reproduce every figure are at the GitHub repo \href{https://github.com/mtrencseni/network-effects-revisited}{mtrencseni/network-effects-revisited}.

\section{Bias across graphs and sizes}

This section compares the five graph families in the linear model. To compare graphs, one has to choose what to hold fixed. One choice is $\nu\bar k$, the peer effect felt by a user with the average number of friends. The other is $\nu\rho(A)$, the distance from the stability limit $\nu\rho = 1$. On the homogeneous families the two are the same, because $\rho(A) \approx \bar k$. On the heavy-tailed families $\rho(A)$  is several times $\bar k$, so at the same $\nu\rho$  the peer effect per friend is several times smaller. Most figures fix $\nu\rho$: on each graph, $\nu$  is set to $x/\rho(A)$, so that $\nu\rho(A) = x$, for values of $x$  from 0.1 to 0.99. Every family is then at the same distance from instability, and the results can be plotted against $\nu\rho$  on one axis. A difference between families at a fixed $\nu\rho$  then comes partly from their different $\nu$, not only from how the network responds to the same user behavior. Figures \ref{fig:figM_1d_attenuation_nukbar} and \ref{fig:figM_2b_size_fixed_nukbar} fix $\nu\bar k$  instead.

\subsection{Stability and amplification}
\begin{figure}[!htbp]\centering
\includegraphics[width=0.9\columnwidth]{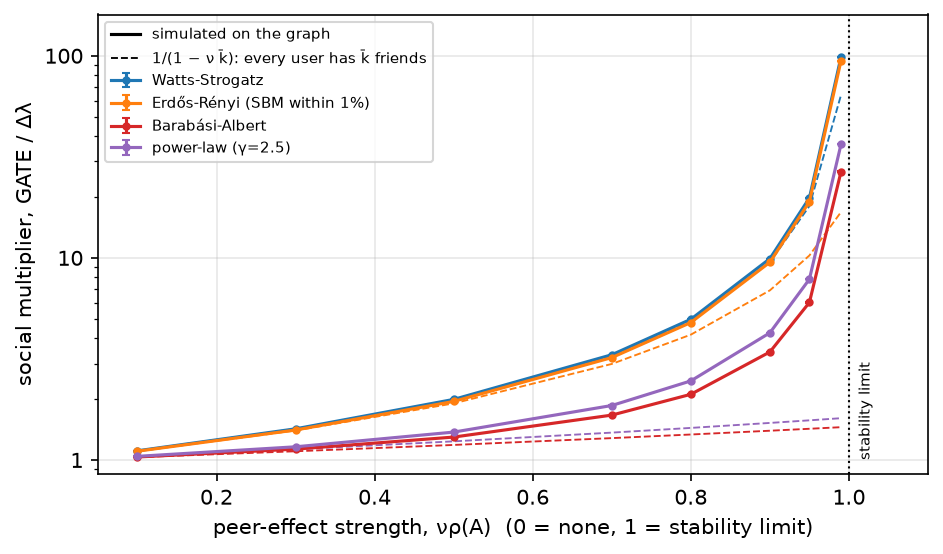}
\caption{The network multiplies the treatment effect, and on every graph family the multiplier grows without bound as $\nu\rho(A)$ approaches 1. Solid: the social multiplier GATE$/\Delta\lambda$ from simulation. Dashed: the prediction from the average number of friends, $1/(1 - \nu\bar k)$, which is close on graphs without hubs and badly wrong on Barab\'asi-Albert and power-law graphs. The Erd\H{o}s-R\'enyi curve also stands for SBM.}\label{fig:figM_1_multiplier}
\end{figure}

A finite steady state exists only when $\nu\rho(A) < 1$  (Section 2), so the stability limit is set by the largest eigenvalue of the graph, not by the average number of friends. The two differ when friend counts are unequal. A user with $k_{\max}$  friends alone makes $\rho(A) \geq \sqrt{k_{\max}}$, and more generally $\rho(A) \approx \max\Bigl\{\langle k^2\rangle/\langle k\rangle, \sqrt{k_{\max}}\Bigr\}$  (Appendix B.1); both terms grow with the spread of friend counts. On the homogeneous families $\rho(A)$  is close to $\bar k$; on the heavy-tailed families it is several times larger (Table~\ref{tab:families}). A peer effect that looks safe from the average number of friends can therefore make posting grow without bound.

Figure \ref{fig:figM_1_multiplier} shows the social multiplier $m = \mathrm{GATE}/\Delta\lambda$: the extra content the whole network produces for each unit of extra content that the treatment adds directly. On every family the simulated multiplier (solid) grows slowly at first and diverges as $\nu\rho$  approaches 1. The dashed curves show the prediction from the average number of friends, $1/(1 - \nu\bar k)$, which is the multiplier if every user had exactly $\bar k$  friends and a lower bound on every undirected graph (Mulholland and Smith~\cite{r18}). On each graph $\nu = x/\rho(A)$, so on the horizontal axis the dashed curve is $1/(1 - x\,\bar k/\rho(A))$. On the homogeneous families $\rho(A)$  is slightly larger than $\bar k$, so the dashed curve diverges just past the stability limit, at $x = \rho(A)/\bar k$; it is close to the simulation for small $x$  and falls behind as $x$  approaches 1. On the heavy-tailed families $\rho(A)$  is about three times $\bar k$, so at the stability limit $x = 1$  the dashed curve is only about $1/(1 - 1/3) = 1.5$. A closed form with $\rho(A)$  in place of $\bar k$  follows the simulated curves (Appendix B.1).

\subsection{What the A/B test reports}
\begin{figure}[!htbp]\centering
\includegraphics[width=0.9\columnwidth]{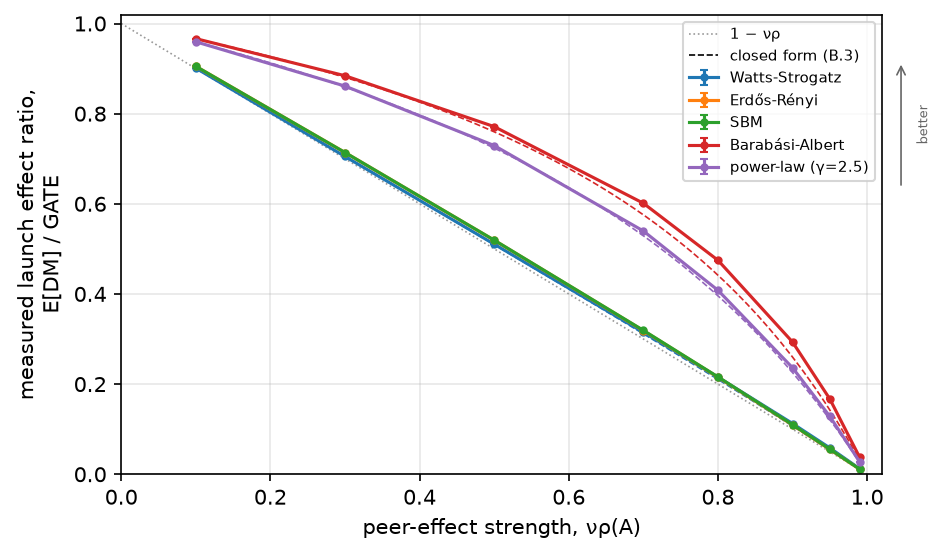}
\caption{An A/B test with 10\% of users treated reports about $1 - \nu\rho$ of the launch effect (dotted) on graphs without hubs; at the same $\nu\rho$, graphs with hubs lie above the line. Dashed: the closed form of Appendix B.3.}\label{fig:figM_1b_attenuation}
\end{figure}
\begin{figure}[!htbp]\centering
\includegraphics[width=0.9\columnwidth]{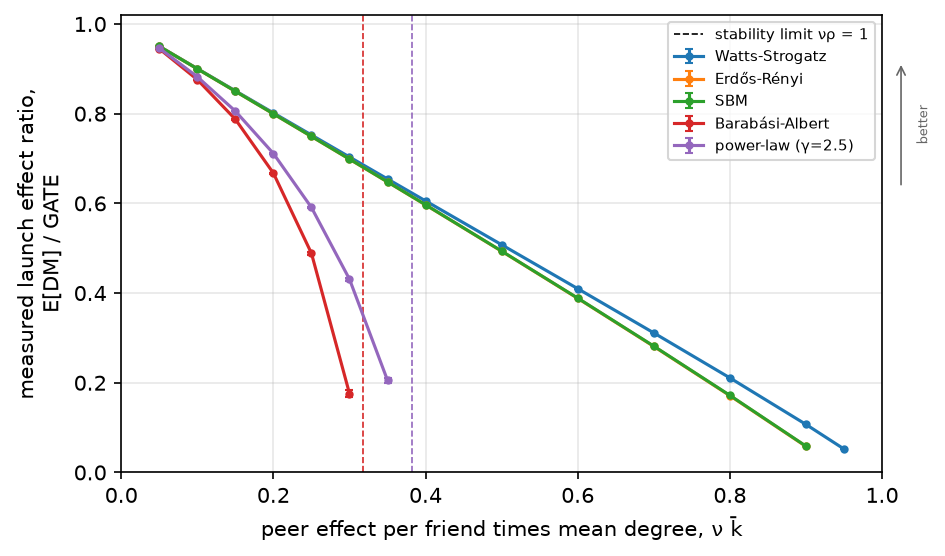}
\caption{At the same peer effect per friend, $\nu\bar k$, the test reports less of the launch effect on graphs with hubs, and almost nothing as they approach their stability limit (dashed vertical lines, where $\nu\rho = 1$). Erd\H{o}s-R\'enyi lies under SBM.}\label{fig:figM_1d_attenuation_nukbar}
\end{figure}
\begin{figure}[!htbp]\centering
\includegraphics[width=0.9\columnwidth]{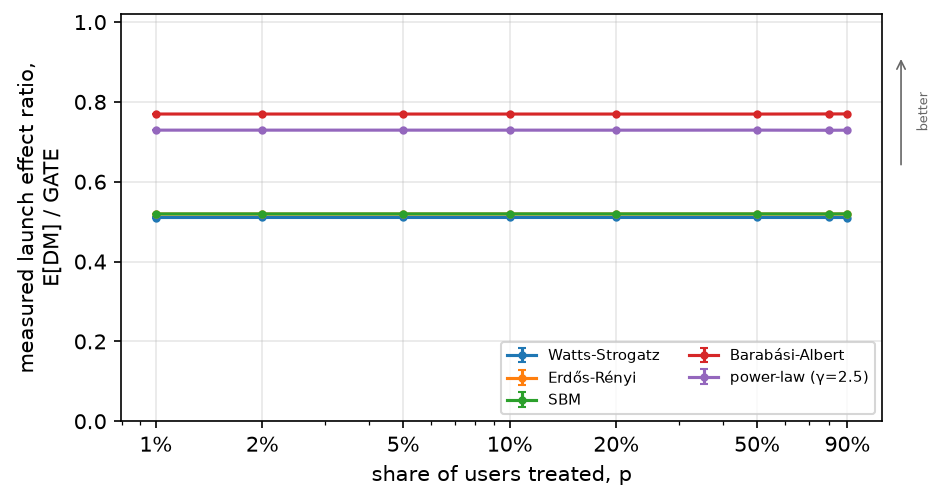}
\caption{The share of the launch effect the test reports does not depend on the share of users treated ($\nu\rho = 0.5$).}\label{fig:figM_1c_treated_fraction}
\end{figure}

To see why the A/B test does not report the GATE, follow the extra content the treatment causes. The model is linear, so the treatment adds $\Delta\lambda Mz$  to production: user $i$  gains $\Delta\lambda M_{ii}$  from their own treatment and $\Delta\lambda M_{ij}$  from each other treated user $j$. Averaged over users, $\Delta\lambda\sum_j M_{ij}$  is the GATE and $\Delta\lambda M_{ii}$  is the ADE (Eqs.~\ref{eq:gate} and \ref{eq:ade}). The gain from everyone else, when everyone else is treated, is therefore $\mathrm{GATE} - \mathrm{ADE}$  on average.

With a share $p$  of users treated at random, each other user is treated with probability $p$, whether or not user $i$  is. On average, then:

\begin{itemize}
\item a control user gains $p(\mathrm{GATE} - \mathrm{ADE})$  from the treated users around them;
\item a treated user gains the same, plus $\mathrm{ADE}$  from their own treatment;
\item after a full launch, every user gains the GATE.
\end{itemize}

Weighted by the arm sizes, the gains add up to $p\,\mathrm{GATE}$, so the mean production $\bar c(p)$  with a share $p$  treated rises in a straight line with the treated fraction, from no one treated to the full launch:
\begin{equation}E[\bar c(p)] = \bar c(0) + p\,\mathrm{GATE}\label{eq:linear}\end{equation}
The term $p(\mathrm{GATE} - \mathrm{ADE})$  is the same in both arms, because being treated does not change whether one's friends are treated. It cancels in the DM, the difference of the two arms, so $E[\mathrm{DM}] = \mathrm{ADE}$  for every $p$, up to terms of order $1/N$  (Eq.~\ref{eq:edm}). The test reports the direct effect, including the small part of the network effect that returns to the treated user, and misses the rest. Since the DM is the treated gain minus the control gain, what it misses is
\begin{equation}\begin{aligned}&\mathrm{GATE} - E[\mathrm{DM}]\\ &\quad= \mathrm{GATE} - (\text{treated gain} - \text{control gain})\\ &\quad= (\mathrm{GATE} - \text{treated gain}) + \text{control gain}\\ &\quad= (1 - p)(\mathrm{GATE} - \mathrm{ADE}) + p(\mathrm{GATE} - \mathrm{ADE})\end{aligned}\label{eq:gap}\end{equation}
The first term is the missing indirect effect: treated users gain less than after a launch, because only a share $p$  of their friends are treated. The second is the spillover into control, which the DM subtracts. With few users treated, the gap is almost all the first term (Appendix F.3).

How large is the gap? By Eq.~\ref{eq:gate} the GATE is $\Delta\lambda\,\mathbf{1}^{\top}M\mathbf{1}/N$, with $M = I + \nu A + (\nu A)^2 + \dots$. For large $t$, $(\nu A)^t \approx (\nu\rho)^t vv^{\top}$, with $v$  the Perron vector (Section 2), so the long paths add up to about $vv^{\top}/(1 - \nu\rho)$: the extra content of every treated user reaches every other user. On a graph where all users have about the same number of friends, $\mathrm{GATE} \approx \Delta\lambda/(1 - \nu\rho)$. By Eq.~\ref{eq:ade} the ADE is $\Delta\lambda\operatorname{tr}(M)/N$, and the trace keeps only the content that returns to the treated user. The growing part $vv^{\top}/(1 - \nu\rho)$  adds only $1/(1 - \nu\rho)$  to the trace, spread over all $N$  users, which is small until $1 - \nu\rho$  is of order $1/N$  (Appendix B.3); the rest comes from short round trips, so $\mathrm{ADE} \approx \Delta\lambda$. The test therefore reports about $1 - \nu\rho$  of the launch effect: half at $\nu\rho = 0.5$, a tenth at $\nu\rho = 0.9$. Figure \ref{fig:figM_1b_attenuation} shows that the homogeneous families follow this line (dotted). The heavy-tailed families lie above it, because at the same $\nu\rho$  their peer effect per friend is smaller. With the multiplier $m \approx 1 + \nu\bar k/(1 - \nu\rho)$  of Appendix B.1, the test reports about $1/m = (1 - \nu\rho)/(1 - \nu\rho + \nu\bar k)$  of the launch effect; this is $1 - \nu\rho$  when $\rho \approx \bar k$, and larger when $\rho$  is several times $\bar k$. The dashed curves are the closed form of Appendix B.3.

At the same peer effect per friend, the order reverses (Figure \ref{fig:figM_1d_attenuation_nukbar}). The heavy-tailed families now lie below the homogeneous ones, and their share falls toward zero as $\nu\bar k$  approaches $\bar k/\rho(A)$, where $\nu\rho$  reaches 1. The same user behavior is amplified more on a graph with hubs, and the test misses more of it. As the derivation predicts, the share does not depend on the treated fraction (Figure \ref{fig:figM_1c_treated_fraction}).

\subsection{Network size}
\begin{figure}[!htbp]\centering
\includegraphics[width=0.9\columnwidth]{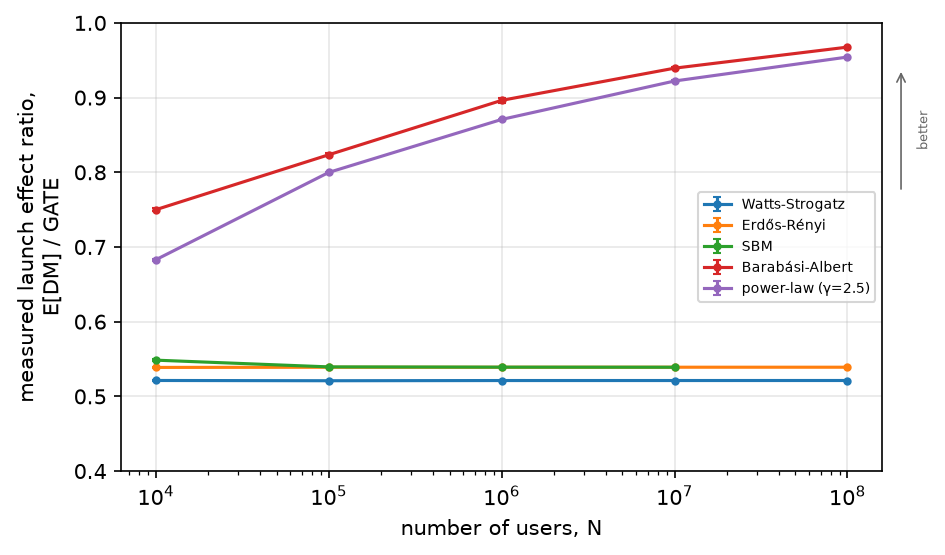}
\caption{On graphs without hubs the test reports the same share of the launch effect at 10 thousand and at 100 million users. On graphs with hubs the share rises, because at a fixed $\nu\rho = 0.5$ the peer effect per friend falls as the graph grows. $\bar k = 10$; SBM up to $10^7$ users.}\label{fig:figM_2_size_fixed_nurho}
\end{figure}
\begin{figure}[!htbp]\centering
\includegraphics[width=0.9\columnwidth]{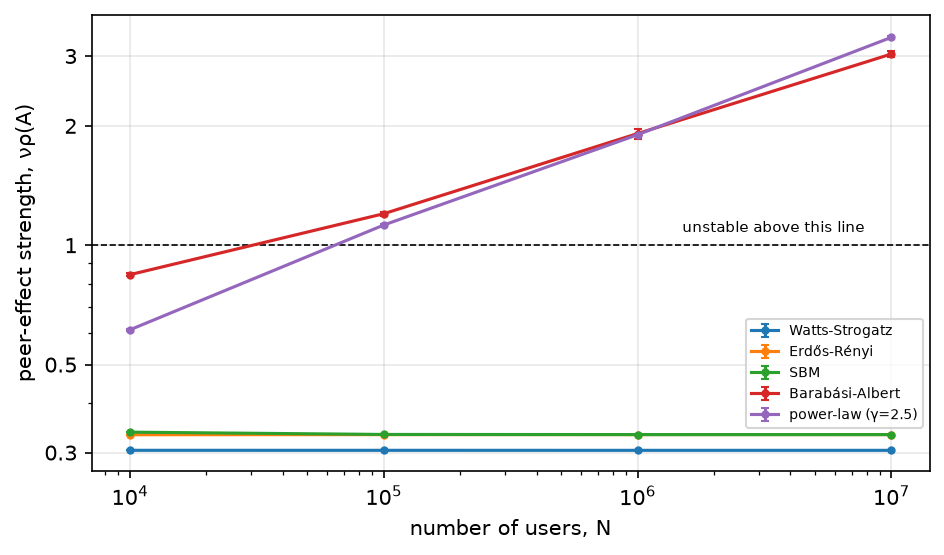}
\caption{On graphs with hubs, a peer effect per friend that is stable on a small network is unstable on a larger one: at a fixed $\nu\bar k = 0.3$, $\nu\rho(A)$ grows with the number of users and crosses the stability limit (dashed). $\bar k = 10$.}\label{fig:figM_2b_size_fixed_nukbar}
\end{figure}

Figure \ref{fig:figM_2_size_fixed_nurho} shows the share of the launch effect the test reports at a fixed $\nu\rho = 0.5$  as the network grows from $10^4$  to $10^8$  users. On the homogeneous families it does not change: a test on a small network reports the same share as on a large one. On the heavy-tailed families it rises toward 1. This does not make the test more accurate on these graphs. Their largest degrees, and with them $\rho(A)$, grow with $N$  while $\bar k$  stays fixed, so a fixed $\nu\rho$  means a smaller peer effect per friend on a larger graph, and there is less network effect to miss (Appendix F.4).

Figure \ref{fig:figM_2b_size_fixed_nukbar} shows the other side, at a fixed $\nu\bar k = 0.3$. On the homogeneous families $\nu\rho$  is the same at every size. On the heavy-tailed families it grows with $N$  and crosses 1: with degree exponent $\gamma = 2.5$  and the cutoff $k_{\max} = \sqrt N$, $\rho$  grows like $N^{1/4}$. A peer effect that is stable on a small network can be unstable on a larger one with the same average number of friends.

\section{Experiment duration}
\begin{figure}[!htbp]\centering
\includegraphics[width=0.9\columnwidth]{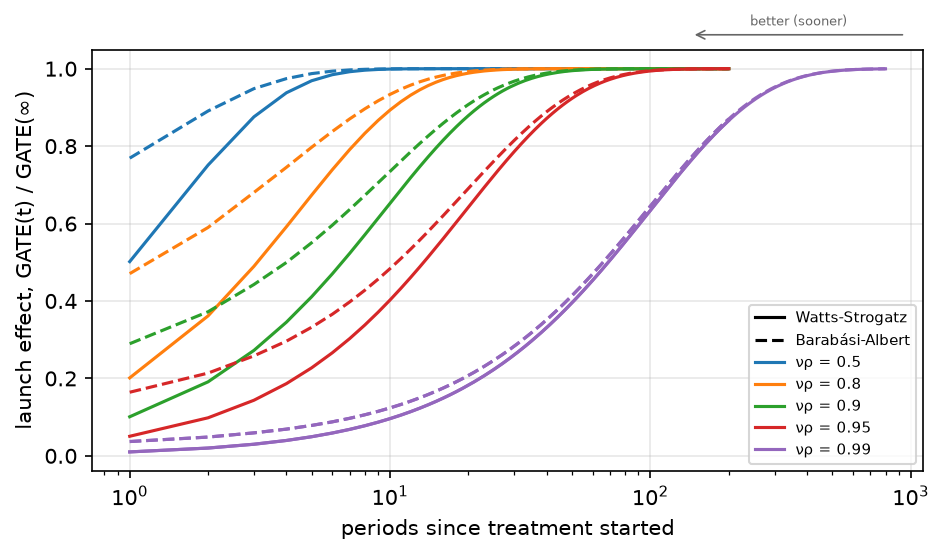}
\caption{After launch, the launch effect keeps building up for many periods, the longer the closer $\nu\rho$ is to 1. GATE(t) relative to its limit, on Watts-Strogatz (solid) and Barab\'asi-Albert (dashed) graphs.}\label{fig:figM_3_gate_paths}
\end{figure}
\begin{figure}[!htbp]\centering
\includegraphics[width=0.9\columnwidth]{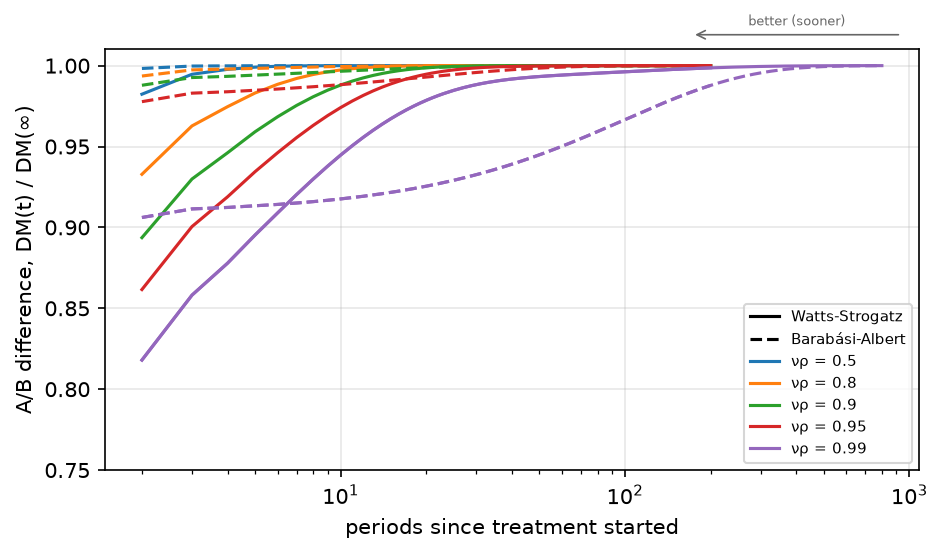}
\caption{The A/B difference settles within the first periods, except very close to the stability limit, so running the test longer does not close the gap. DM(t) relative to its limit, on the same graphs.}\label{fig:figM_3b_dm_paths}
\end{figure}

The network starts in its untreated steady state, and treatment starts at $t = 0$. Each period the extra content travels one more hop, so the launch effect builds up as a geometric sum, $\mathrm{GATE}(t) = \Delta\lambda\,\mathbf{1}^{\top}(I + \nu A + \dots + (\nu A)^{t-1})\mathbf{1}/N$, one term per hop (Eckles, Karrer and Ugander~\cite{r8}, eq. 11, give the row-normalized version). The terms shrink like $(\nu\rho)^t$, so the time to approach the limit grows like $1/(1 - \nu\rho)$: the closer $\nu\rho$  is to 1, the more terms in the series matter, the slower the convergence (Figure \ref{fig:figM_3_gate_paths}).

The A/B difference settles much sooner (Figure \ref{fig:figM_3b_dm_paths}). The direct effect appears in the first period. The later terms, the content passed on through friends, reach both arms alike and mostly cancel in the DM, as in Section 4.2. The DM is therefore close to its final value while the launch effect is still growing, and running the test longer does not reduce the gap. The settling times do not change with graph size on the homogeneous families (Appendix E). One period is one hop of influence; Section 6 takes it to be one day. At moderate $\nu\rho$, stopping at the first significant daily check gives too many false positives, as without network effects (Johari et al.~\cite{r14}).

\section{Designs and estimators}

Sections 4 and 5 show that an A/B test with a user-level split reports only part of the launch effect GATE, and that running it longer does not help. There are two ways to reduce the gap: change the design, so that a user's friends are more often in the same arm (cluster randomization, 6.1), or change the estimator, so that it targets the GATE instead of the direct effect (6.2). Near the stability limit the test also loses power, and variance reduction recovers only part of it (6.3).

Unlike Sections 4 and 5, the simulations of this section include the randomness of a real A/B test. Four sources of randomness enter: the assignment of the treatment group and the Poisson observation noise of Section 2, drawn anew for each simulated test; and the graph itself and each user's intrinsic content production, drawn anew for each graph, the latter in Sections 6.1 and 6.2 from a gamma distribution with mean 1 instead of being 1 for everyone. Errors and power are computed over all four.

\subsection{Cluster randomization}
\begin{figure}[!htbp]\centering
\includegraphics[width=0.9\columnwidth]{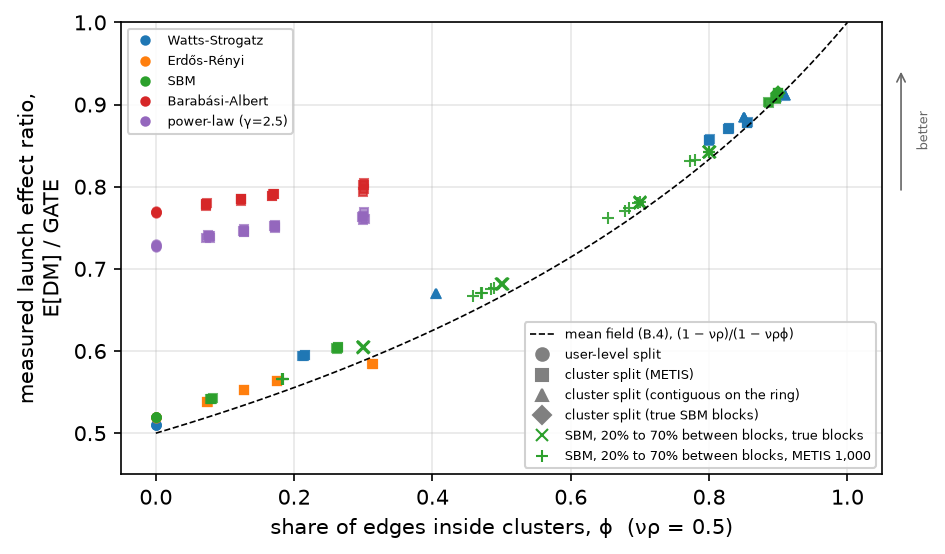}
\caption{Treating whole clusters of friends together brings the test closer to the launch effect, the more so the larger the share $\phi$ of friendships kept inside clusters ($\nu\rho = 0.5$). Dashed: the mean field of Eq.~\ref{eq:cluster}, which holds on graphs without hubs. Crosses and plus signs: SBM graphs with 20\% to 70\% of edges between blocks, with the true blocks and with METIS clusters of 1,000.}\label{fig:figM_4_clusters_phi}
\end{figure}
\begin{figure}[!htbp]\centering
\includegraphics[width=0.9\columnwidth]{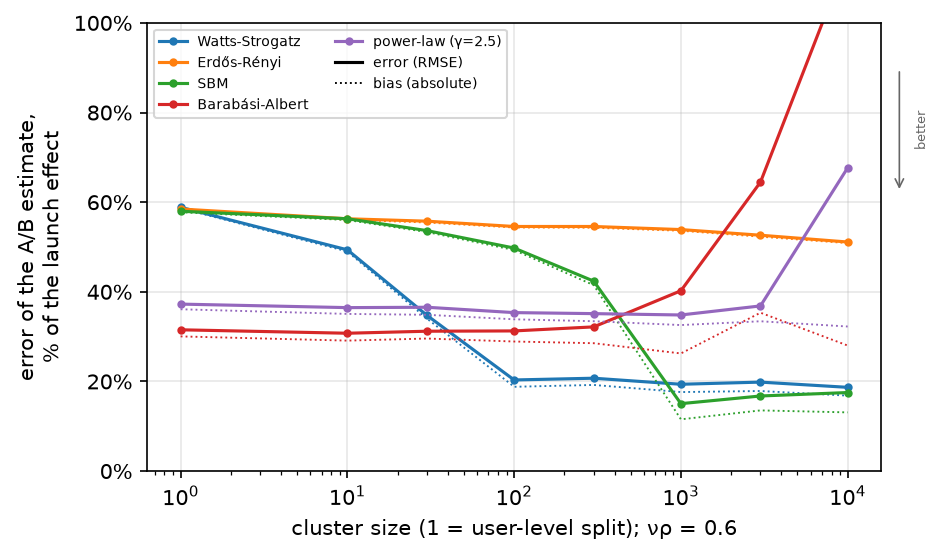}
\caption{Clusters cut the error on graphs with communities or local clustering, barely help on Erd\H{o}s-R\'enyi graphs, and with very large clusters make it worse on graphs with hubs. Error (RMSE, solid) and bias (dotted) of the DM relative to the GATE against cluster size, $\nu\rho = 0.6$, weekly Poisson counts.}\label{fig:figM_4b_clusters_size}
\end{figure}

The test misses the gain that treated users would get from treated friends (Section 4.2). Treating whole clusters of connected users puts more of a user's friends in the same arm, so treated users get more of that gain and control users less. Let $\phi$  be the share of friendships whose two users are in the same cluster, as defined by the clustering algorithm used to assign treatment in the A/B test, and so are always in the same arm. It depends on both the graph and the algorithm. In a user-level split each user is their own cluster, so by definition $\phi = 0$. To see how much clusters help, take every user to be average: a share $\phi$  of their friends are in their own cluster, hence in their own arm, the rest are in other clusters, each treated with probability $p$, and the peer influence on each user is $\nu\rho$. The mean extra production $x_T$  of a treated user and $x_C$  of a control user then satisfy
\begin{equation}\begin{aligned}x_T &= \Delta\lambda + \nu\rho\,[\phi x_T + (1 - \phi)(p\,x_T + (1 - p)\,x_C)]\\ x_C &= \nu\rho\,[\phi x_C + (1 - \phi)(p\,x_T + (1 - p)\,x_C)]\end{aligned}\label{eq:meanfield}\end{equation}
The term for friends in other clusters is the same in both arms and cancels in the difference, as the term $p(\mathrm{GATE} - \mathrm{ADE})$  does in Section 4.2, so the result does not depend on $p$:
\begin{equation}\begin{aligned}x_T - x_C &= \Delta\lambda + \nu\rho\phi\,(x_T - x_C)\\ E[\mathrm{DM}] = x_T - x_C &= \Delta\lambda/(1 - \nu\rho\phi)\end{aligned}\label{eq:dmcluster}\end{equation}
After a full launch every user gains $x = \Delta\lambda + \nu\rho x$, so $\mathrm{GATE} = \Delta\lambda/(1 - \nu\rho)$, and the share of the launch effect the test reports is (Appendix B.4 gives a more accurate version)
\begin{equation}E[\mathrm{DM}]/\mathrm{GATE} \approx (1 - \nu\rho)/(1 - \nu\rho\phi)\label{eq:cluster}\end{equation}
At $\phi = 0$  this is the $1 - \nu\rho$  of Section 4.2, and it rises to 1 as $\phi$  approaches 1. Figure \ref{fig:figM_4_clusters_phi} shows that the homogeneous families follow it (dashed). Clusters help only as far as the graph has structure to keep inside them. The true blocks of SBM graphs and contiguous stretches of the Watts-Strogatz ring keep most edges inside, and the test then reports most of the launch effect. The heavy-tailed families lie above the curve, already at $\phi = 0$, because the mean field gives every user the peer influence $\nu\rho$, while a typical user's is closer to $\nu\bar k$  (Section 4.2). Clusters move these points little, because on graphs without communities METIS (Karypis and Kumar~\cite{r15}) keeps few friendships inside clusters; the same holds for Erdős-Rényi graphs.

Clusters reduce the bias but add variance, because with fewer and larger units the two arms differ more by chance (Figure \ref{fig:figM_4b_clusters_size}). They cut the total error on graphs with communities or local clustering, barely help on Erdős-Rényi graphs, and on graphs with hubs large clusters make it worse (Appendix F.8).

\subsection{Estimators of the launch effect}
\begin{figure*}[!tp]\centering
\includegraphics[width=\textwidth]{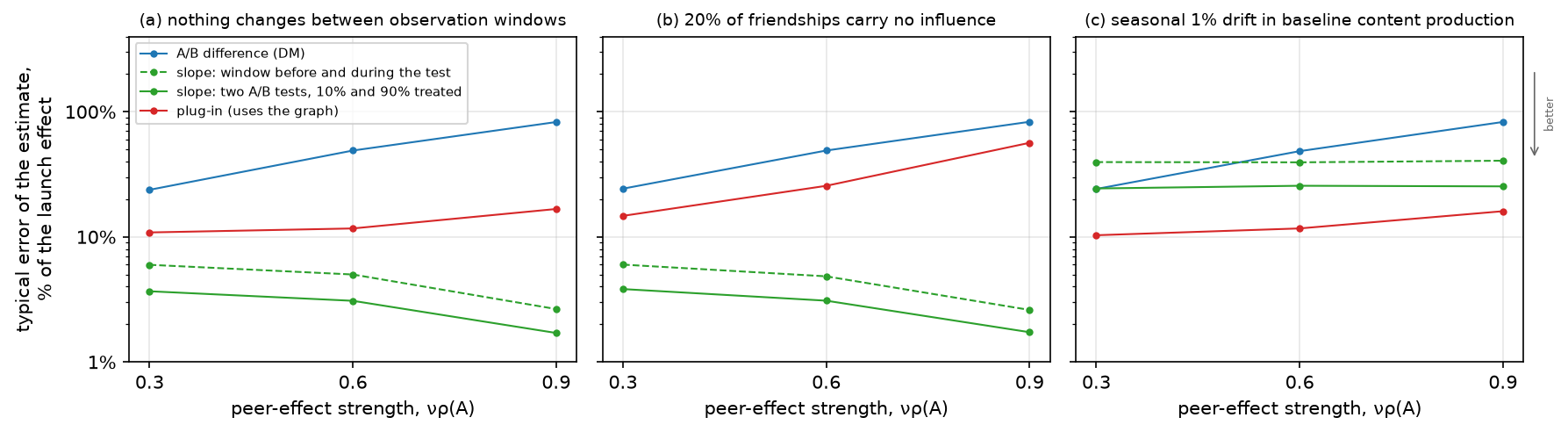}
\caption{Two simple estimators recover the launch effect when the observation windows are comparable; with a small drift in baseline posting the slope is far off, and with wrong friendships the plug-in is far off. Typical error (RMSE) of the DM, the slope estimator (a window before and one during the test, or two A/B tests at 10\% and 90\% treated) and the plug-in, as a percentage of the launch effect, in three settings (a) to (c) described in the text; mean over the five families.}\label{fig:figM_5_extrapolation}
\end{figure*}

In this section, instead of increasing the ratio $E[\mathrm{DM}]/\mathrm{GATE}$, we look at two approaches that measure the launch effect GATE indirectly. The first compares the whole population in two observation windows with different treated fractions. Per Eq.~\ref{eq:linear} from Section 4.2, mean content production rises in a straight line with the treated fraction, and the slope of that line is the GATE. Two windows give two points on the line, and the slope between them estimates the GATE without using the graph or the model. The two points can be a window before the test, with no one treated, and a window during the test (the historical baseline of Yu et al.~\cite{r25}), or two A/B tests with 10\% and 90\% treated (a staggered rollout; Cortez et al.~\cite{r6}). The second approach, the plug-in, uses the friendship graph and every user's content production, treated and control, once it has settled following the start of the test. It fits the linear model of Eq.~\ref{eq:model} to the test: each user's production is regressed on their own binary treatment variable $z_i$, which estimates $\Delta\lambda$, and on the total production of their friends, which estimates $\nu$. A plain regression overstates $\nu$, because friends raise each other's production and active users tend to be friends with each other; two-stage least squares uses only the part of the friends' production that their random treatment causes (Kelejian and Prucha~\cite{r16}; Bramoullé et al.~\cite{r4}). The GATE then follows from Eq.~\ref{eq:gate} with the estimates and the graph, $\mathrm{GATE} = \widehat{\Delta\lambda}\,\mathbf{1}^{\top}(I - \hat\nu A)^{-1}\mathbf{1}/N$.

Figure \ref{fig:figM_5_extrapolation} shows how far each estimate typically is from the true launch effect, as a percentage of it: at 10\% an estimate is typically off by a tenth of the GATE, at 100\% by as much as the GATE itself. The panels change one condition at a time. (a) When nothing changes between the two windows, both estimators are close to the GATE, the slope closer than the plug-in, while the DM is far from it. (b) When a fifth of the recorded friendships carry no influence, and as many real influence links are missing from the record, the plug-in is far off; the slope does not use the graph. (c) When everyone's baseline posting changes between the windows, for example with the season, the slope is far off: the treatment changes mean production by only a share $\Delta p\,\Delta\lambda/\lambda$, where $\Delta p$  is the difference in treated fraction, so even a small drift is a large fraction of the effect. The plug-in uses one window and is not affected. The slope also needs production to have settled in each window (Section 5) and mean production to be linear in the treated fraction, which the threshold model of Section 8.1 is not.

The estimators of this section rely on assumptions to estimate the launch effect GATE, which is the quantity experimenters want to know. The A/B test needs no such assumptions, because both arms are observed in the same window, so a drift affects both and cancels. But it measures a different quantity: the difference in means between the treated and the control arm, which on a social network is only part of the GATE (Sections 4 and 5). Figure \ref{fig:figM_5_extrapolation} shows that the estimators of this section measure the GATE accurately when their assumptions hold and break down when they do not. The A/B test always measures the DM reliably, but the DM is not the GATE.

\subsection{Power near the stability limit and CUPED}

\begin{figure}[!htbp]\centering
\includegraphics[width=0.9\columnwidth]{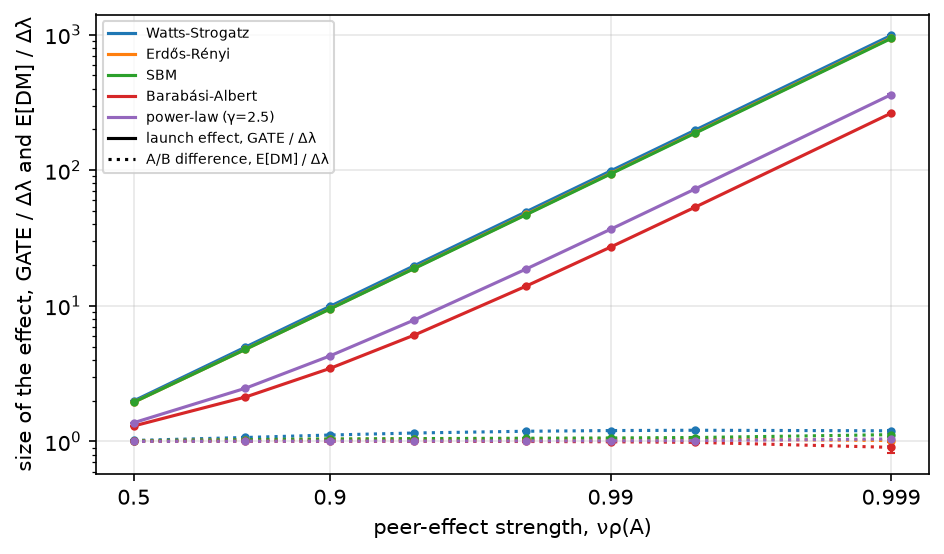}
\caption{As $\nu\rho$ approaches 1 the launch effect (solid) grows without bound, while the A/B difference (dotted) stays near the direct effect $\Delta\lambda$. The axis is logarithmic in the distance from the stability limit: each tick is ten times closer to 1.}\label{fig:figM_6_critical_effects}
\end{figure}

Near the stability limit the launch effect GATE grows without bound, while the A/B difference $E[\mathrm{DM}]$, which equals the ADE up to terms of order $1/N$  (Eq.~\ref{eq:edm}), stays near the direct effect $\Delta\lambda$  in the simulated range, where $1 - \nu\rho \gg 1/N$  (Figure \ref{fig:figM_6_critical_effects}; Section 4.2; Appendix B.3).

\begin{figure}[!htbp]\centering
\includegraphics[width=0.9\columnwidth]{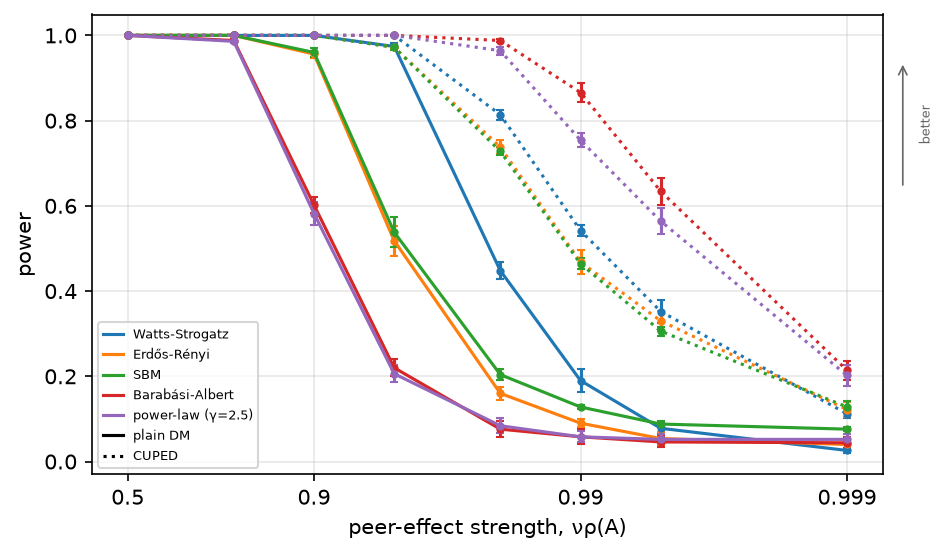}
\caption{The test loses power near the stability limit; pre-period adjustment (CUPED, dotted) recovers part of it when each user's baseline is the same before and during the test. One-sided test, $\Delta\lambda/\lambda = 0.05$. The axis is logarithmic in the distance from the stability limit.}\label{fig:figM_6b_critical_power}
\end{figure}

Near the stability limit the test also loses power: its signal $E[\mathrm{DM}]$, approximately the ADE, stays near $\Delta\lambda$, while production itself, and the differences between users that make the test noisy, grow in proportion to the GATE. As $\nu\rho$  approaches 1 the power falls toward the significance level (Figure \ref{fig:figM_6b_critical_power}).

A standard way to reduce the variance of an A/B test is CUPED (Deng et al.~\cite{r7}): from each user's outcome it subtracts a multiple of the same user's outcome before the test, with the multiple chosen to minimize the variance. Assignment is random, so the adjustment has mean zero: CUPED estimates the same $E[\mathrm{DM}]$  as the plain test, with less noise, and the gap to the GATE stays the same in expectation. Here every user has the same intrinsic production $\lambda$, but untreated production still differs between users because of their position in the graph: users with more friends see more content and post more. These differences persist from before the test, so CUPED removes them. It cannot remove the randomness of the observed counts: the posts a user is observed to make in a week are a Poisson count around their production rate, drawn anew each week (the observation noise of Section 2). It recovers part of the lost power when there is no seasonality, that is, when each user's production rate does not change between the two windows (Figure \ref{fig:figM_6b_critical_power}, dotted).

\section{Treatment that changes the graph}

So far we modeled treatments that only increase content production, by $\Delta\lambda$. In this section we examine treatments that change the graph itself: friend recommendations add edges, and invitations add users.

\subsection{Treatment adds edges}
\begin{figure}[!htbp]\centering
\includegraphics[width=0.9\columnwidth]{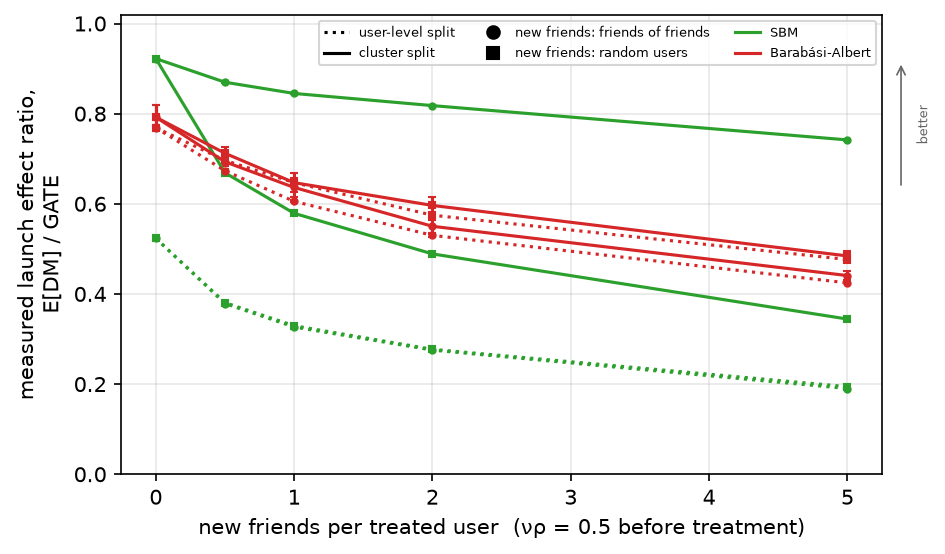}
\caption{When the feature makes users add friends, the test reports a shrinking share of the launch effect as more friends are added. A cluster split (solid) keeps most of the effect only when the new friends are inside the cluster (friends of friends, circles); with random users (squares) its share falls as fast as under the user-level split. Dotted: user-level split. SBM (green) and Barab\'asi-Albert (red) graphs. SBM and Barab\'asi-Albert stand for graphs with and without communities; the other families are described in Appendix F.14.}\label{fig:figM_7_new_edges}
\end{figure}

Here each treated user gets the intrinsic boost $\Delta\lambda$  and also forms Poisson($r$) new undirected edges, with $r = 0.5$  to 5, at $\nu\rho = 0.5$  on the original graph. Two variants are simulated separately: in one, each new friend is a friend of a friend, as with typical friend recommendations; in the other, a uniformly random user. The GATE compares everyone getting both with no one getting either. A full launch makes the whole network denser, and a denser network amplifies every user's intrinsic production $\lambda$, not only $\Delta\lambda$. The launch effect therefore grows quickly with $r$  and soon far exceeds the effect of $\Delta\lambda$  alone. In an A/B test the new edges also reach control users, so both arms receive part of the densification, and that part cancels in the DM. Figure \ref{fig:figM_7_new_edges} shows the result. Under a user-level split (dotted) the share of the launch effect the test reports falls as $r$  grows, to a lower level on homogeneous graphs than on heavy-tailed ones. A cluster split (solid), with clusters computed on the graph before the test, reports more than the user-level split in both variants, but keeps most of the effect only when the new edges stay inside treated clusters. On SBM graphs they do when the new friends are friends of friends, who are mostly in the same cluster. When they are random users, about half of whom are in the other arm, the share falls as fast as under the user-level split.

\subsection{Treatment adds users}
\begin{figure}[!htbp]\centering
\includegraphics[width=0.9\columnwidth]{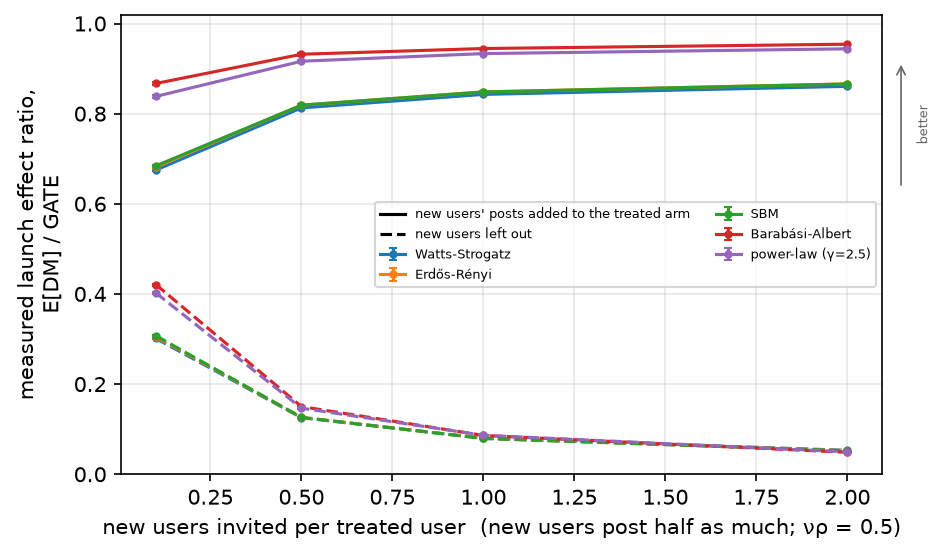}
\caption{When the feature makes users invite new users, adding the new users' posts to the treated arm's total per original user (solid) captures the new users' own posts, which a launch adds too, and misses only the network effect, as in Section 4.2; the more users are invited, the larger the share it recovers. Leaving them out (dashed) captures only the treatment's effect on existing users. New users post half as much as existing users and befriend the inviter and two draws from the inviter's friends.}\label{fig:figM_7b_new_users}
\end{figure}

Here each treated user gets the intrinsic boost $\Delta\lambda$  and invites Poisson($r$) new users; a new user connects to the inviter and to $q$  draws from the inviter's friends and produces $a\lambda$, with $r = 0.1$  to 2, $q = 0$  or 2 and $a = 0.25$  to 1 at $\nu\rho = 0.5$. The new users exist only in the launched world, so the launch effect is measured as the change in total production per original user. The per-user effect, averaged over all users including the new ones, can be negative while the total is positive, because new users post less. The A/B test's estimate adds the new users' production to the total of the inviter's arm, which is always the treated arm, and divides by the number of original users in that arm:
\begin{equation}\widehat{\mathrm{DM}} = \big(\sum_{i \in T} c_i + \sum_{j \in \mathrm{new}} c_j\big)/N_T - \sum_{i \in C} c_i/N_C\label{eq:newusers}\end{equation}
with $T$  and $C$  the original treated and control users. This estimate (Figure \ref{fig:figM_7b_new_users}, solid) captures the new users' own posts, which a launch adds too: only treated users invite, so it counts about $r$  new users per original treated user, as many per original user as after a launch. It misses the network effect, as in Section 4.2: the treatment's boost and the new users' posts also reach control users, who are friends of treated users and of new users. The more users are invited, the larger the part of the GATE that is the new users' own posts, and the larger the share the estimate recovers; the share is also larger on heavy-tailed graphs, where the network effect is smaller at the same $\nu\rho$. When the new users are left out (dashed), the estimate reports a small share of the GATE, falling as $r$  grows: it measures the effect on existing users only. The ordinary difference of per-user means, which also counts the new users in the treated denominator, can have the wrong sign, and is negative in almost every setting simulated (Appendix F.15). A three-type mean field gives the GATE and the estimates (Appendix B.6).

\section{Nonlinear responses}

\subsection{Threshold contagion}
\begin{figure}[!htbp]\centering
\includegraphics[width=0.9\columnwidth]{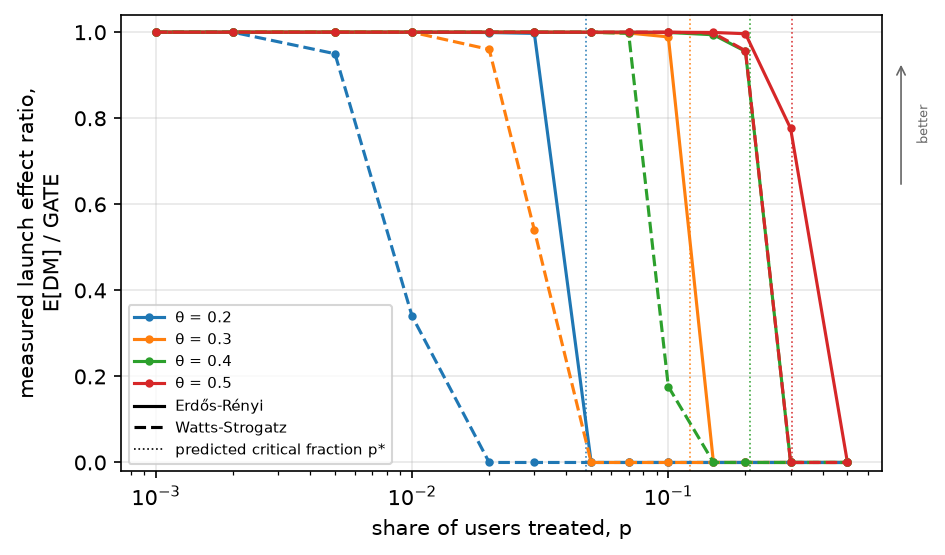}
\caption{A behavior that spreads once enough friends adopt it: the test reports almost all of the launch effect below a critical treated fraction, and almost nothing above it. One color per threshold $\theta$; solid: Erd\H{o}s-R\'enyi, where the tree formula of Appendix B.7 predicts the critical fraction (dotted vertical lines); dashed: Watts-Strogatz, where friends are often friends of each other and the switch comes earlier. The other families are described in Appendix F.16.}\label{fig:figM_9_threshold}
\end{figure}

Some features are useful only when friends use them too, such as a group chat or a multiplayer game. A user adopts such a feature once enough of their friends have adopted it. Adoption can then spread: each new user brings some of their friends over their threshold, and those bring their own friends. Here the treatment's effect does not fade with distance as in the linear model. A control user enough of whose friends adopt the feature adopts it as well, and behaves exactly like a treated user. As the spread reaches the control arm, the A/B test increasingly compares two groups that both use the feature: all of the treated users, and a growing share of the control users. We model this with the fractional threshold model of Watts~\cite{r23}: a node becomes active when it is treated or when a fraction $\theta$  of its neighbors are active, and an active node produces $\lambda + \Delta\lambda$, treated or not. There is no other peer effect: the only way users influence each other is by making each other active.

Every treated user is active, and a control user is active only if the spread has reached them, so $\mathrm{DM} = \Delta\lambda\,(1 - P(\text{control user active}))$. After a full launch every user is active, so $\mathrm{GATE} = \Delta\lambda$, and
\begin{equation}\mathrm{DM}/\mathrm{GATE} = 1 - P(\text{control user active})\label{eq:threshold}\end{equation}
The test misses exactly the share of control users the spread has reached. Figure \ref{fig:figM_9_threshold} shows what this means. Below a critical treated fraction $p^*$  the spread stalls, few control users are reached, and the test reports almost all of the launch effect. Above $p^*$  the spread reaches the whole graph, control users included, and the test reports almost nothing, exactly when the feature is spreading to everyone. The switch is sharp, and $p^*$  is higher the higher the threshold $\theta$. A formula that assumes a user's friends are not friends of each other predicts $p^*$  from the degree distribution alone (Gleeson and Cahalane~\cite{r10}; Appendix B.7). It holds on Erdős-Rényi, Barabási-Albert and power-law graphs. On Watts-Strogatz graphs, where friends are often friends of each other, a few treated users in one neighborhood push each other over the threshold, so the switch comes earlier and is less sharp. On SBM graphs the treated share varies between blocks, and a block that starts above $p^*$  can start the spread early (Appendix F.16).

Mean production does not rise in a straight line with the treated fraction but jumps at $p^*$, so the slope estimators of Section 6.2 do not apply: two tests on either side of $p^*$  compare a stalled spread with a full one. What remains is a warning sign: an A/B difference that falls as the treated fraction grows means the behavior is reaching the control arm.

\subsection{Concave peer response}
\begin{figure}[!htbp]\centering
\includegraphics[width=0.9\columnwidth]{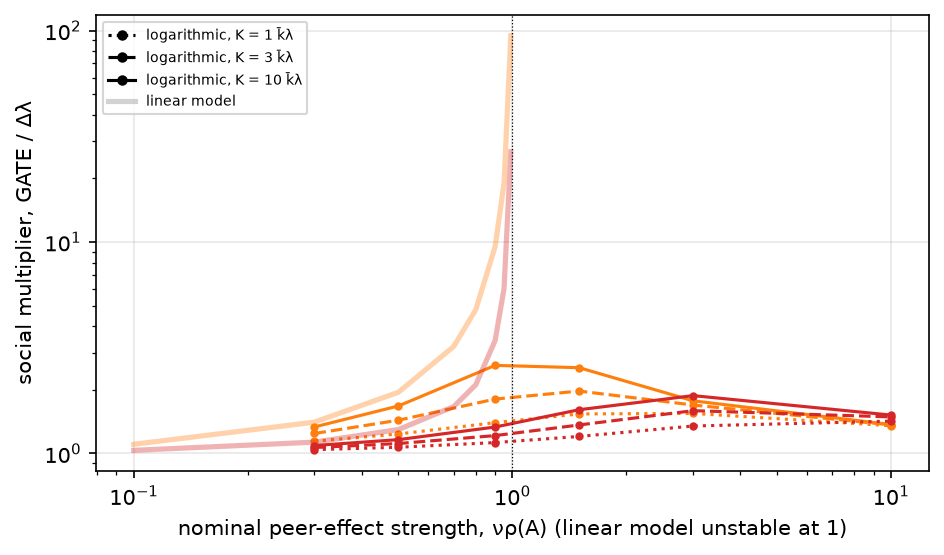}
\caption{If each additional post a user sees adds less than the previous one, posting never grows without bound. Social multiplier with the response $g(s) = K\log(1 + s/K)$, $K = \sigma\bar k\lambda$, on Erd\H{o}s-R\'enyi (orange) and Barab\'asi-Albert (red) graphs for $\sigma = 1$, 3, 10 (dotted, dashed, solid), against the linear model, which diverges at $\nu\rho = 1$.}\label{fig:figM_8_concave_multiplier}
\end{figure}
\begin{figure}[!htbp]\centering
\includegraphics[width=0.9\columnwidth]{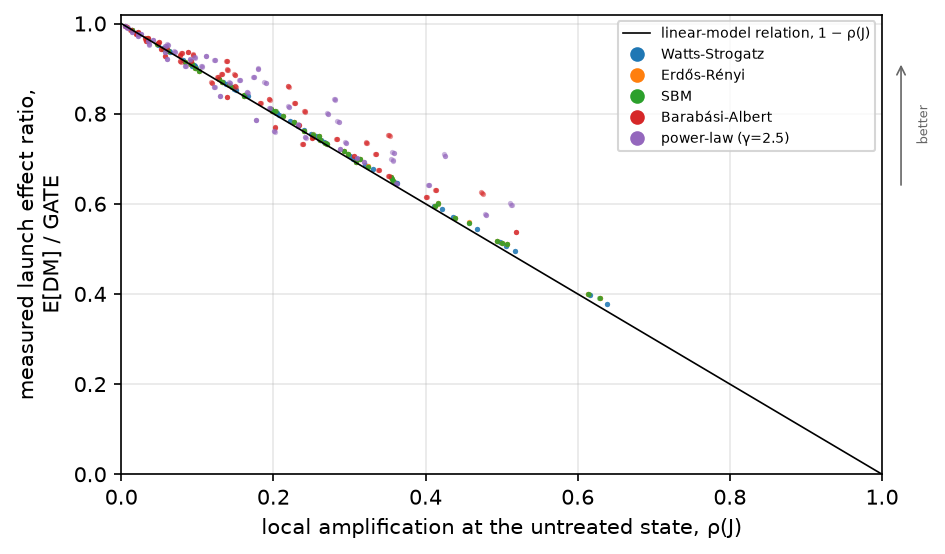}
\caption{With a concave response the test reports about $1 - \rho(J)$ of the launch effect, as in the linear model with $\rho(J)$ in place of $\nu\rho$. Every configuration of the grid: both responses, all families, 10\% treated.}\label{fig:figM_8b_concave_attenuation}
\end{figure}

In the linear model each additional unit of content a user sees adds the same amount to the user's production. Here we check what changes when each additional unit adds less than the previous one. This section replaces the sum $Ac$  in Eq.~\ref{eq:model} by a concave function of it, $c = \lambda + \Delta\lambda z + \nu g(Ac)$, with $g(s) = K\log(1 + s/K)$, linear for $s \ll K$  and logarithmic beyond,\footnote{The factor $K$  in front of the logarithm makes $g'(0) = 1$  and $g(s) \leq s$, so $\nu$  keeps its meaning from the linear model and $K$  sets only where the response bends. The saturating form also has $g'(0) = 1$.} and with the saturating $g(s) = s/(1 + s/K)$  as a check (Appendix F.17). The saturation scale is $K = \sigma\bar k\lambda$  with $\sigma = 0.3$, 1, 3 and 10; $K \to \infty$  is the linear model. The peer effect is given by its nominal value $\nu\rho(A)$  from 0.3 to 10, the stability parameter of the linear model.

With the concave response posting never grows without bound (Figure \ref{fig:figM_8_concave_multiplier}). With a positive intercept and $g(s)/s \to 0$, the model has exactly one steady state for every $\nu$  and every graph, and iteration from $c = \lambda$  converges to it (Kennan~\cite{r17}; Belhaj, Bramoullé and Deroïan~\cite{r3}). Every simulated configuration converged, including far past $\nu\rho = 1$. The instability of Sections 4.1 and 4.3 is a property of the linear response.

A small treatment still acts on the network almost as in the linear model. Around the untreated steady state it responds to the matrix $J = \nu\,\mathrm{diag}(g'(s))A$, $s = Ac$, in which each user's peer effect is scaled by the slope of $g$  at the user's own exposure. The users who see the most content, here the high-degree users, are damped most. The local amplification $\rho(J)$  takes the place of $\nu\rho$, and it is always below 1. Here it is computed from the known response; estimating $J$  from data was not studied. On the grid simulated here $\rho(J)$  stays well below 1, so the network effect stays moderate. This is a property of the grid, not of concave responses in general: as $K$  grows the model approaches the linear one.

Figure \ref{fig:figM_8b_concave_attenuation} shows that the A/B test follows the linear relation with $\nu\rho$  replaced by $\rho(J)$: the points lie near the line $1 - \rho(J)$, close to it on the homogeneous families and with more scatter on the heavy-tailed ones. As in the linear model, the share does not depend on the treated fraction. Cluster randomization behaves as in the linear model with $\rho(J)$  in place of $\nu\rho$, and the slope estimators of Section 6.2 stay close to the GATE (Appendix F.17); the plug-in was not run with the concave response.

\section{Practical guidance}

This section lists specific advice on how to estimate the full launch effect GATE, in the order of the work, for an experimenter who has a real social network and can run simulations on it.

\begin{enumerate}[itemsep=1pt,topsep=2pt,leftmargin=*]
\item \textbf{Interpret the DM as the direct effect (ADE), not the launch effect (GATE).} The effect through friends reaches treated and control users alike and cancels in the difference, so the test misses most of it. Neither a longer test nor a different treated fraction changes this (Figures \ref{fig:figM_3b_dm_paths} and \ref{fig:figM_1c_treated_fraction}), so choose the test length and the treated fraction for statistical power alone. Near the stability limit the test loses power, and CUPED recovers part of the lost power when users' baselines are stable (Section 6.3).
\item \textbf{Measure the graph.} Compare its largest eigenvalue $\rho(A)$  with the mean degree $\bar k$: if $\rho(A)$  is several times $\bar k$, a few very connected users amplify the network effect, and the test misses more than the average number of friends suggests (Figure \ref{fig:figM_1d_attenuation_nukbar}). Partition the graph and measure the share $\phi$  of friendships inside clusters: it indicates how much a cluster design can help (Eq.~\ref{eq:cluster}).
\item \textbf{Estimate the peer effect $\nu$  from past tests,} with the regression of Section 6.2 on the graph of who actually sees whose content, for example logged impressions. If production flattens as users see more content, the linear fit does not apply; fitting the concave model of Section 8.2 was not studied here.
\item \textbf{Treat features that change the network separately.} For growth features, add the new users' posts to the inviter's arm (Eq.~\ref{eq:newusers}). For features that add friendships, expect the test to miss more the more friendships are added (Figure \ref{fig:figM_7_new_edges}). Some features spread on their own once enough of a user's friends have adopted them (Section 8.1). For these, run the test at several treated fractions. If the A/B difference gets smaller as more users are treated, the feature is spreading to the control group, and the test underestimates its effect.
\item \textbf{Simulate the next test before running it,} over a range of $\nu$  around the estimate, and choose the design with the smallest error against the simulated launch effect: a user-level split, clusters, or two treated fractions. Clusters help only where the graph has communities or local clustering, and must be analyzed by cluster.
\item \textbf{Estimate the launch effect directly when the conditions allow:} two observation windows with different treated fractions give the GATE without the graph, if production has settled, baseline production is stable between the windows, and mean production is linear in the treated fraction, which spreading features are not (Sections 6.2 and 8.1).
\end{enumerate}

\section{Conclusion}

In the linear model a finite steady state exists only when $\nu\rho(A) < 1$. On heavy-tailed graphs, with a few highly connected users, the largest eigenvalue $\rho(A)$  is several times the mean number of friends $\bar k$, so the network effect is much larger than $\bar k$  suggests. An A/B test reports the direct effect of a treatment: about a $1 - \nu\rho(A)$  fraction of the launch effect on graphs with homogeneous degrees, and about $(1 - \nu\rho(A))/(1 - \nu\rho(A) + \nu\bar k)$  on heavy-tailed graphs; running longer does not close the gap. Cluster randomization closes more of it the larger the share of edges kept inside clusters; the slope estimator closes it without the graph when baseline production is stable. Treatment that changes the graph, or behavior that spreads once enough friends adopt it, can make the test miss most of the launch effect. With a concave response these relations hold approximately, with $\rho(J)$  taking the role of $\nu\rho(A)$, where $\rho(J) < 1$  is the largest eigenvalue of the matrix $J$  which describes how a small treatment spreads.

\flushcolsend
\clearpage
\appendix
\renewcommand{\thefigure}{\arabic{figure}}\renewcommand{\thetable}{\arabic{table}}
\section*{Appendices}
\footnotesize
\renewcommand{\dbltopfraction}{0.95}\renewcommand{\dblfloatpagefraction}{0.8}\setcounter{dbltopnumber}{3}\setlength{\dblfloatsep}{10pt plus 2pt minus 2pt}

\subsection*{A. Limitations}

The main limitations of this work are the following.

\textbf{Random graphs.} The graphs are random graph families, not real social networks. The results depend on graph structure, such as communities and very connected users, so on a real network the share of the launch effect a test reports, and how much clusters help, should be checked by simulating on that network (Section 9).

\textbf{Noise model.} The noise model is the mildest one: independent variation in each user's observed counts. If each user's posting is itself random, the randomness spreads to their friends and is several times larger near the stability limit, so real tests may have less power than Figure \ref{fig:figM_6b_critical_power} shows.

\textbf{Graph growth.} Treatment that adds edges or users is evaluated at its final state, after all new friendships and users are in place. A test read while the network is still growing sees less of the launch effect than Section 7 reports.

\subsection*{B. Closed forms and approximations}
\begin{figure*}[!b]\centering
\includegraphics[width=\textwidth]{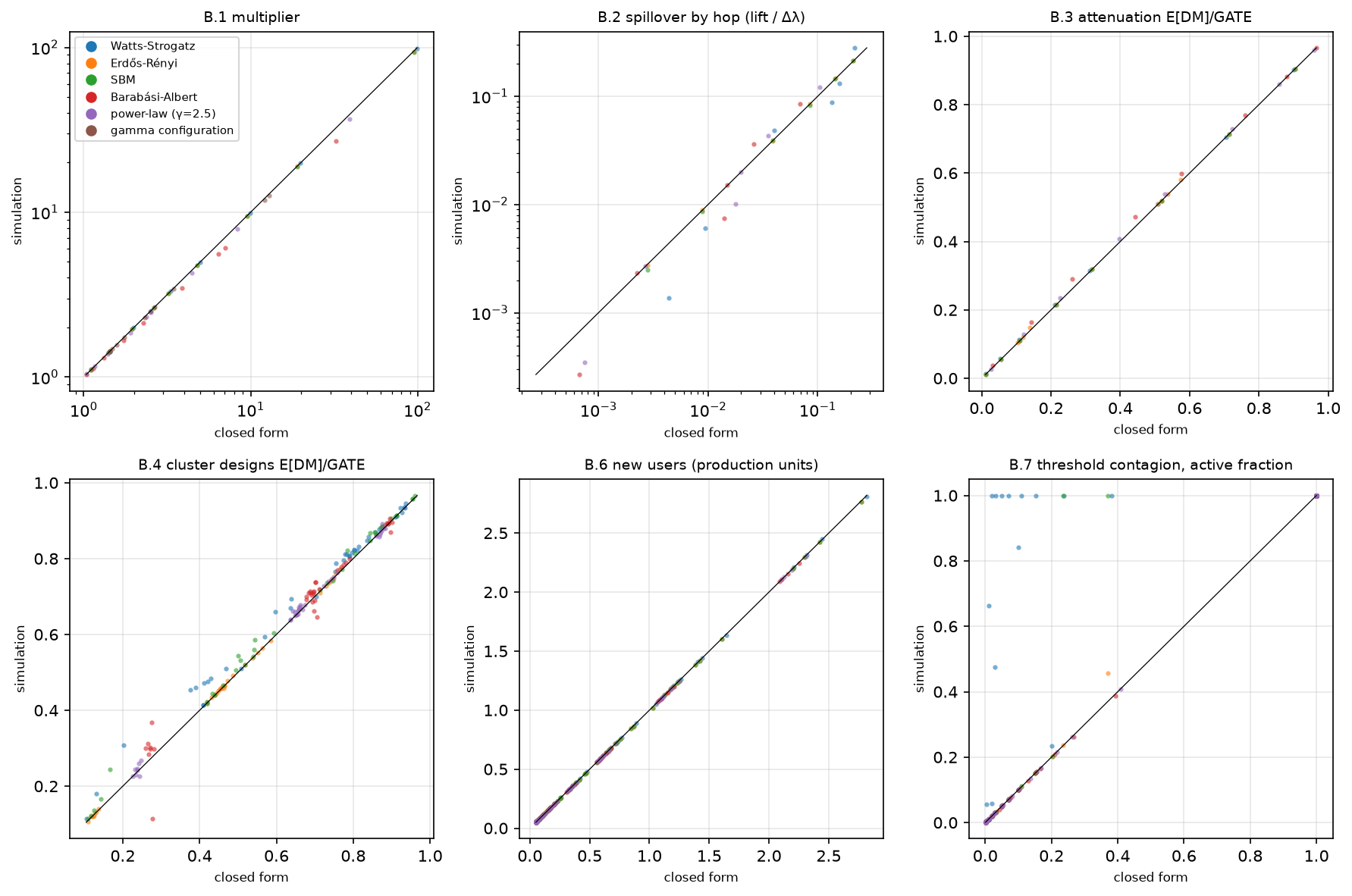}
\caption{The closed forms of Appendix B (horizontal) against the simulation (vertical): points on the diagonal mean that the formula predicts the simulated value. Six panels, for B.1 to B.4, B.6 and B.7. The main misses are Watts-Strogatz cascades (B.7) and hop distances (B.2), and Barab\'asi-Albert cluster designs near the stability limit (B.4).}\label{fig:figA_1_closed_forms}
\end{figure*}

Notation: $M = (I - \nu A)^{-1}$; $m = \mathbf{1}^{\top}M\mathbf{1}/N$, the multiplier; $\tau = \operatorname{tr}(M)/N$; $\kappa = \langle k^2\rangle/\langle k\rangle$, the mean degree of a neighbor; $b = \kappa - 1$, the mean excess degree (Newman, Strogatz and Watts~\cite{r19}). The relations marked exact hold for every graph under the stated conditions. The others are approximations, with their assumptions and failures stated and no error bound; Figure \ref{fig:figA_1_closed_forms} compares B.1 to B.4, B.6 and B.7 with the simulation. Some inputs are read off the graph ($\bar k$, $\langle k^2\rangle$, $k_{\max}$, the degree histogram, the triangles per node, $\phi$); others come from the solved model: the spectral radius $\rho$, the spread of $M\mathbf{1}$  across nodes and the GATE. The exact identities and the homogeneous-degree limits were derived twice, independently; the degree-weighted versions used for the heavy-tailed graphs were derived once and checked only against the simulation. The code is in \texttt{netfx/theory.py}.

\textbf{Exact relations.} These restate, for the model of Section 2, the linear-model results of Eckles, Karrer and Ugander~\cite[eqs. 6 to 9 and 11]{r8} and Hu, Li and Wager~\cite[Example 2 and Theorem 1]{r12}, with the finite-population correction for the difference in means added; their Theorem 1 gives $\mathrm{ADE} + \mathrm{AIE}$  as the effect of a marginal increase in the treatment probability, which equals the GATE here because $E[\bar c(p)]$  is linear in $p$. Under Bernoulli assignment, $E[\mathrm{DM}] = (N\,\mathrm{ADE} - \mathrm{GATE})/(N - 1)$. Under cluster assignment with $C$  equal-size clusters treated with probability 1/2, $E[\mathrm{DM}] = \Delta\lambda(C\sigma - m)/(C - 1)$  with $\sigma = \sum_c \mathbf{1}_c^{\top}M\mathbf{1}_c/N$; for unequal clusters, such as METIS partitions, the relation holds to first order in the spread of cluster sizes. Under a treated fraction $p$, $E[\bar c(p)] = \bar c(0) + p\,\mathrm{GATE}$. After launch, $\mathrm{GATE}(t) = \Delta\lambda\sum_{s<t}\mathbf{1}^{\top}(\nu A)^s\mathbf{1}/N$. In the threshold model, $\mathrm{DM}/\mathrm{GATE} = 1 - P(\text{control node active})$.

\textbf{B.1 Multiplier and spectral radius.} $\rho(A) \approx \max\Bigl\{\kappa, \sqrt{k_{\max}}\Bigr\}$. For expected-degree graphs this holds when the two terms are far apart; for $G(n, p)$  the leading-order value is $\max\Bigl\{np, \sqrt{k_{\max}}\Bigr\}$; for configuration models it is the heuristic of Castellano and Pastor-Satorras~\cite{r5}. On the power-law graph used here the two terms are within a factor of 2, no theorem applies, and the formula is an empirical approximation. The multiplier $m \approx 1 + \nu\bar k/(1 - \nu\rho)$  follows from the rank-one approximation $A \approx kk^{\top}/(N\bar k)$, the expected adjacency matrix of the Chung-Lu model, with the measured $\rho$  at the pole. With gamma-distributed degrees, $\kappa = \bar k(1 + cv^2)$, so the annealed approximation predicts stability when $cv < \sqrt{1/(\nu\bar k) - 1}$, with $cv$  the coefficient of variation of the degree distribution. The value $1/(1 - \nu\bar k)$  is a lower bound for undirected graphs, since $\mathbf{1}^{\top}A^t\mathbf{1}/N \geq \bar k^t$. Accuracy: close on every family away from the limit; near the limit B.1 overshoots on the heavy-tailed graphs, where the Perron vector is concentrated on the hubs.

\textbf{B.2 Green's function of a locally tree-like graph.} With $g = (1 - \sqrt{1 - 4b\nu^2})/(2b)$  and $T = 1/(1 - bg)$, $M_{ii} = 1/(1 - k_i g)$  and $M_{ij} = M_{jj}(\nu T)^d$  for nodes at distance $d$. For a random $k$-regular graph $M_{ii}$  is the Kesten-McKay value; for a degree distribution, one cavity value $g$  with branching factor $b$  is the effective-medium approximation of the exact tree equations. The compared $\tau$  includes the leading non-tree closed walks, $2t\nu^3T^3M_{ii}$  with $t$  the triangles per node, which matters only on Watts-Strogatz graphs. Accuracy: close on Erdős-Rényi and SBM graphs, less so on heavy-tailed graphs at large $\nu\rho$. B.2 fails with short cycles: on Watts-Strogatz graphs a node's neighbors share most of their neighbors, so the tree overcounts the nodes at distance 2.

\textbf{B.3 Attenuation.} $E[\mathrm{DM}]/\mathrm{GATE} = (N\tau - m)/((N - 1)m)$, with $\tau = E_k[1/(1 - kg)]$  plus the triangle term of B.2 and $m$  from B.1. For a random $k$-regular graph the ratio tends to $(1 - \nu k)/(1 - kg) = (1 - \nu k)(1 + k\nu^2 + (2k^2 - k)\nu^4 + \dots)$, so the correction to $1 - \nu\rho$  shrinks like $1/\bar k$. Near the limit, on a connected graph with unit Perron vector $v$, $M \approx vv^{\top}/(1 - \nu\rho)$, so $E[\mathrm{DM}] = \Delta\lambda(N - (\mathbf{1}^{\top}v)^2)/(N(N - 1)(1 - \nu\rho)) + O(1)$. The pole term is of order $1/(N(1 - \nu\rho))$, so $E[\mathrm{DM}]$  stays near its bulk value until $1 - \nu\rho$  is of order $1/N$. On a connected regular graph $v = \mathbf{1}/\sqrt N$, the Perron term cancels and $E[\mathrm{DM}]$  has a finite limit. Accuracy: close on every family away from the limit; near it the formula is low on Watts-Strogatz graphs, where cycles longer than triangles raise $\tau$, and on Barabási-Albert graphs, where B.1 overshoots $m$. With the measured $\tau$  and $m$  the exact relation reproduces the simulation on every family.

\textbf{B.4 Cluster designs.} Mean field, for homogeneous degrees. Every user is taken to be average: a share $\phi$  of their friends is in their own cluster, hence in the same arm, the rest are in other clusters, each treated with probability $p$, and the peer influence on each user is $\nu\rho$. The mean extra production $x_T$  of treated and $x_C$  of control users then satisfies $x_T = \Delta\lambda + \nu\rho[\phi x_T + (1 - \phi)(p x_T + (1 - p) x_C)]$  and $x_C = \nu\rho[\phi x_C + (1 - \phi)(p x_T + (1 - p) x_C)]$. The cross-cluster term cancels in the difference, so $E[\mathrm{DM}] = x_T - x_C = \Delta\lambda/(1 - \nu\rho\phi)$, and with $\mathrm{GATE} = \Delta\lambda/(1 - \nu\rho)$, $E[\mathrm{DM}]/\mathrm{GATE} \approx (1 - \nu\rho)/(1 - \nu\rho\,\phi_C)$, where $\phi_C = (C\phi - 1)/(C - 1)$  corrects $\phi$  for $C$  clusters and tends to $\phi$  for many clusters. The assumption $\mathrm{GATE} = \Delta\lambda/(1 - \nu\rho)$  fails on heavy-tailed graphs (Section 6.1). To first order in $\nu\rho$  this is $1 - \nu\rho(1 - \phi)$: the relative bias is the spillover strength times the share of cut edges, the bias measure of Viviano et al.~\cite[Lemma 3.1]{r22}. Tree version: with $r = \phi_C$, a walk of length $d$  ends in its start cluster with probability $1/C + (1 - 1/C)r^d$, so the within-cluster mass is $\sigma = m/C + (1 - 1/C)F$  with $F = E_k[M_{ii}(1 + kr\nu T/(1 - br\nu T))]$  (B.2), and the exact relation gives $E[\mathrm{DM}]/\mathrm{GATE} = F/m$. With $\sigma_\lambda^2$  the variance of intrinsic production across users and $\bar c$  the mean production, the variance over assignments is $\mathrm{Var}(\mathrm{DM}) \approx (4/N)[\sigma_\lambda^2\,\omega + \mathrm{Var}_i((M\mathbf{1})_i) + \bar c/W + 2\nu^2\,\mathrm{Var}(k)\,\phi\,\bar c^2/(1 - \nu\bar k\phi_C)^2]$, with $\omega = \sum_c\mathbf{1}_c^{\top}M^2\mathbf{1}_c/N$, evaluated with the measured GATE and spread of $M\mathbf{1}$; the last term is a heuristic for the shared degree fluctuation of a cluster. Accuracy: the tree version of the bias is close on every family at moderate $\nu\rho$, and low near the limit on the clustered and community graphs; the mean-field version is slightly low and misses on heavy-tailed graphs. The standard deviation is close on the homogeneous families and much too small on heavy-tailed graphs with large clusters, where the hubs, linked to each other, fall into a few clusters.

\textbf{B.5 Power.} The expected test statistic is $z_{\mathrm{stat}} \approx (\Delta\lambda/\lambda)(\tau/m)\sqrt N/(2\sqrt{cv_b^2 + 1/(W\lambda m)})$, with $cv_b$  the relative spread of $M\mathbf{1}$  across nodes and $W$  the observation window; $m$  and $cv_b$  are measured on the graph.

\textbf{B.6 Treatment adds users.} A three-type mean field (treated, control, new users) in which a neighbor's production over original edges is weighted by degree as in B.1; a new user links to its inviter and to $q$  of the inviter's neighbors. It gives the total and per-user GATE and the estimates of Section 7.2, close to the simulation in every setting.

\textbf{B.7 Threshold contagion} (Gleeson and Cahalane~\cite{r10}). With $P(k)$  the degree distribution and $u$  the probability that a node reached along an edge becomes active, iterate $u \leftarrow p + (1 - p)\sum_k(kP(k)/\bar k)P[\mathrm{Bin}(k - 1, u) \geq \lceil\theta k\rceil]$  from $u = p$; then $P(\text{active}) = p + (1 - p)\sum_k P(k)P[\mathrm{Bin}(k, u) \geq \lceil\theta k\rceil]$. Accuracy: it places the transition at the simulated point on Erdős-Rényi, Barabási-Albert and power-law graphs; it is late on both SBM and Watts-Strogatz graphs. B.7 fails with clustering and at the threshold itself, where finite graphs are bistable.

\textbf{B.8 Concave response.} For $c = \lambda + \Delta\lambda z + \nu g(Ac)$  with $g$  increasing and concave, $g(0) = 0$  and $g(s)/s \to 0$, the steady state exists and is unique for every $\nu$  and graph (Kennan~\cite{r17}: Corollary 1 for uniqueness, since $F(c) = \lambda + \nu g(Ac)$  is increasing and $F(c) - c$  is strictly ray-concave because $\lambda > 0$  and $g(\theta s) \geq \theta g(s)$; Theorem 2 for existence, with $F(\lambda) \geq \lambda$  and $F(b) < b$  for a large $b$  since $g(s)/s \to 0$; Remark 3 for the convergence of the iteration from $c = \lambda$; Belhaj, Bramoullé and Deroïan~\cite{r3}), and the response to a small treatment is $(I - J)^{-1}\Delta\lambda z$  with $J = \nu\,\mathrm{diag}(g'(Ac))A$. Annealed version: a degree-$k$  node sees $k\theta$, with $\theta = \sum_k kP(k)(\lambda + \nu g(k\theta))/\bar k$; with $w_k = \nu g'(k\theta)$, $m \approx 1 + \langle kw\rangle/(1 - \langle k^2w\rangle/\langle k\rangle)$, and $\rho(J) \approx \langle k^2w\rangle/\langle k\rangle$. For the logarithmic response $kw_k = \nu k/(1 + k\theta/K) < \nu K/\theta$  at every degree, so a node's weight stays bounded as its degree grows, whereas in the linear model it is $\nu k$. Accuracy: for the small treatment, $E[\mathrm{DM}]/\mathrm{GATE} \approx 1/m$  is close to the simulation at every point of the grid. The closed form is a derivative at the untreated state and does not include the change in the slope of $g$  under a finite treatment, so for the large treatment it is somewhat low. The code is \texttt{netfx/concave.py}.

\subsection*{C. Simulator and performance}

The five graph families are built as in the literature (Watts and Strogatz~\cite{r24}; Erdős and Rényi~\cite{r9}; Holland, Laskey and Leinhardt~\cite{r11}; Barabási and Albert~\cite{r2}; the configuration model of Newman, Strogatz and Watts~\cite{r19}), and Appendix F.1 adds a gamma configuration model; cluster designs use METIS partitions (Karypis and Kumar~\cite{r15}). Removing self-loops and duplicate edges lowers the realized mean degree of the power-law graph slightly, and the closed forms use the realized $\bar k$. Table~\ref{tab:setup} lists the setup of every experiment.

\begin{table*}[!tp]
\caption{Setup of each experiment. Unless stated otherwise, $\bar k = 20$, $\lambda  = 1$ and $\Delta \lambda  = 0.05$.}\label{tab:setup}
\centering
\footnotesize
\renewcommand{\arraystretch}{1.2}
\begin{tabularx}{\linewidth}{>{\raggedright\arraybackslash}X>{\raggedright\arraybackslash}X>{\raggedright\arraybackslash}X>{\raggedright\arraybackslash}X>{\raggedright\arraybackslash}X}
\toprule
\textbf{simulation} & \textbf{N and parameters} & \textbf{design} & \textbf{observation} & \textbf{graphs $\times $ assignments} \\
\midrule[\heavyrulewidth]
4.1, 4.2 & $10^{5}$  & Bernoulli 10\% (F.2: 2\%) & exact & 3 to $10 \times  1$ to 20\\
\hline
4.3 size & $10^{4}$ to $10^{8}$ (fixed $\nu \bar k$: to $10^{7}$), $\bar k = 10$  & Bernoulli 10\% & exact & $10 / 10 / 5 / 2 / 1 \times  1$ \\
\hline
4.3 size (F.4) & $10^{4}$ to $10^{9}$, $\bar k = 2$  & Bernoulli 10\% & exact & $1 \times  1$ \\
\hline
5 duration & $10^{5}$  & Bernoulli 10\% & exact, per period & $5 \times  1$ \\
\hline
5 daily checks & $10^{5}$, $\Delta \lambda  = 0$ or 0.002 & Bernoulli 50\% & Poisson, daily, 60 days & $5 \times  40$ \\
\hline
6.1 clustering & $10^{5}$  & Bernoulli or clusters, 50\% & exact & $5 \times  20$ \\
\hline
6.1, 6.2, F.9 to F.11 & $10^{5}$, $\lambda $ \textasciitilde{} Gamma(2, 1/2) & Bernoulli or clusters, 50\% & Poisson, W = 7 & $5 \times  20$ to 50\\
\hline
6.3 near the limit & $10^{5}$, $\Delta \lambda  = 0.05$ or 0 & Bernoulli 50\% & Poisson, W = 7 & $5 \times  100$ \\
\hline
F.6 saturation & $10^{5}$  & Bernoulli or targeted & exact & $5 \times  1$ \\
\hline
F.13 inference & $10^{5}$, $\lambda $ \textasciitilde{} Gamma(2, $1/2$), $\Delta \lambda  = 0$ or 0.002 & Bernoulli or clusters & Poisson, W = 7 & $5 \times  100$ \\
\hline
7 graph change & $10^{5}$  & Bernoulli or clusters, 50\% & exact & $5 \times  10$ \\
\hline
8.1 threshold & $10^{5}$, $\Delta \lambda  = 0.5$  & Bernoulli & exact & $5 \times  20$ \\
\hline
8.2 concave & $10^{5}$, $\Delta \lambda  = 0.05$ and 0.5 & Bernoulli or clusters & exact & $3 \times  3$ to 10\\
\hline
A production noise & $2\cdot 10^{4}$, $\Delta \lambda  = 0$  & none & 2,000 periods & $5 \times  1$ \\
\hline
E size check & $10^{5}$, $10^{6}$, $10^{7}$  & as in 5 to 8 & as in 5 to 8 & $3 / 3 / 1 \times  1$ to 10 \\
\bottomrule
\end{tabularx}
\end{table*}

The simulator stores a graph as two index arrays (compressed sparse rows without a data array) and computes matrix products with multithreaded loops. Steady states are solved by conjugate gradients, many right-hand sides at once, with a check that stops when $\nu\rho \geq 1$. The spectral radius comes from the ARPACK Lanczos solver (SciPy \texttt{eigsh}). At $10^9$  nodes \texttt{eigsh} does not fit in memory next to the graph, so the largest runs use the three-term Lanczos recursion without reorthogonalization, which keeps three vectors; the loss of orthogonality in finite precision adds copies of converged eigenvalues but does not change the largest one. The recursion agrees with ARPACK to at least ten digits on graphs with up to $10^8$  nodes. Graph generators are implemented to scale to $10^9$  nodes.

\subsection*{D. Reproduction of Trencséni~\cite{r20}}

As a validation, the simulator reproduces every number of the notebook of Trencséni~\cite[arXiv:2312.01607v1]{r20} to within $10^{-13}$, and its figures pixel for pixel.

Trencséni~\cite{r20} named several effects of this model. Table~\ref{tab:names} maps them to standard terms and to the closed forms of Appendix B. Here $\phi$  is the share of edges inside the clusters of a cluster design (Section 6.1), $p$  the treated fraction and $cv$  the coefficient of variation of the degree distribution.

\begin{table*}[!tp]
\caption{The effects of the model, their standard names, their closed forms (Appendix B) and their names in Trencséni~\cite{r20}.}\label{tab:names}
\centering
\footnotesize
\renewcommand{\arraystretch}{1.2}
\begin{tabularx}{\linewidth}{>{\raggedright\arraybackslash}X>{\raggedright\arraybackslash}X>{\raggedright\arraybackslash}X>{\raggedright\arraybackslash}X}
\toprule
\textbf{effect} & \textbf{standard name} & \textbf{metric / closed form} & \textbf{Trencséni~\cite{r20} name} \\
\midrule[\heavyrulewidth]
neighbors raise each other's production & peer effect $(\nu )$; social multiplier & multiplier $m \approx  1 + \nu  \bar k/(1 - \nu  \rho )$ (B.1) & network effect\\
\hline
wider degree distribution raises production at fixed $\bar k$  & degree heterogeneity effect & $\rho  \approx  ⟨k^{2}⟩/⟨k⟩ = \bar k(1 + cv^{2})$ (B.1) & degree distribution effect\\
\hline
treatment raises control's production & spillover; indirect effect & lift by hop distance (B.2) & spillover effect\\
\hline
treated nodes miss the boost of treated neighbors & missing indirect effect on the treated, as a share of the GATE; $\approx $ (1 $- p$) AIE/GATE with AIE = GATE $-$ ADE (Hu, Li and Wager~\cite{r12}) & $1 -$ lift of treated / GATE = (1 $-$ p)(1 $- E[\mathrm{DM}]/\mathrm{GATE}$) (F.3) & intrinsic dampening\\
\hline
DM underestimates the GATE & interference bias; attenuation ratio & $E[\mathrm{DM}]/\mathrm{GATE} = (N τ - m)/((N - 1) m)$ (B.3) & experiment dampening\\
\hline
clustered treatment reduces the bias & graph cluster randomization & $E[\mathrm{DM}]/\mathrm{GATE} \approx $ (1 $- \nu  \rho )/(1 - \nu  \rho  φ$) (B.4, homogeneous degrees) & clustering effect\\
\hline
treated fraction & saturation & E[DM] independent of saturation & experiment size effect \\
\bottomrule
\end{tabularx}
\end{table*}

\subsection*{E. Size check}

The main quantities of Sections 5 to 8 were repeated at $N = 10^5$, $10^6$  and $10^7$  (3, 3 and 1 graphs); the concave response of Section 8.2 was run at $N = 10^5$  only. On the homogeneous families every quantity is the same at all three sizes, with one exception: on SBM graphs the 100 blocks grow with $N$, so at the larger sizes clusters of 1,000 cut through the blocks and lose their advantage. On the heavy-tailed families the values drift with $N$  for the reason of Section 4.3: at fixed $\nu\rho$, $\rho$  grows with $N$, so the peer effect on a typical node weakens.

\subsection*{F. Supplementary results}

This appendix holds the detailed results behind Sections 4 to 8, with their figures.

\subsubsection*{F.1 Degree heterogeneity}
\begin{figure}[!htbp]\centering
\includegraphics[width=0.9\columnwidth]{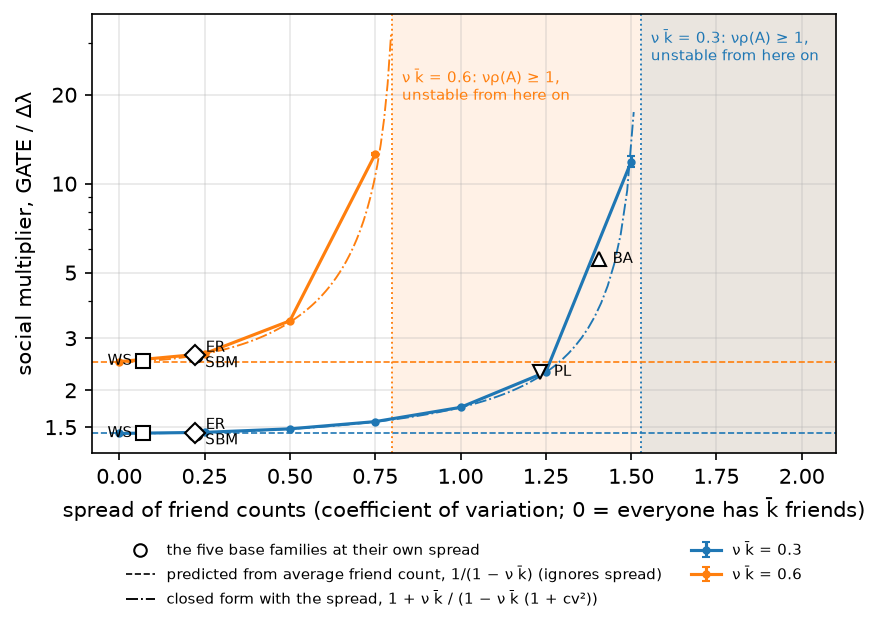}
\caption{At the same average number of friends, the more unequal the friend counts, the larger the multiplier, until production grows without bound (shaded). Gamma configuration model, $\bar k = 20$. Dashed: $1/(1 - \nu\bar k)$, which ignores the spread; dash-dot: the closed form of Appendix B.1, which includes it. Open markers: the five base families at their own spread.}\label{fig:fig5_2_heterogeneity}
\end{figure}

The five base families cannot isolate the effect of the spread of friend counts: on Watts-Strogatz graphs rewiring changes the spread and the clustering together, on Erdős-Rényi and SBM graphs the spread is fixed, and on the heavy-tailed graphs it changes with $N$  or with other parameters. The gamma configuration model changes only the spread, at a fixed $\bar k$  (Figure \ref{fig:fig5_2_heterogeneity}). At a fixed $\nu\bar k$  the multiplier rises with the spread, slowly at first and then sharply, until the graph becomes unstable at $cv = \sqrt{1/(\nu\bar k) - 1}$, the boundary of B.1. The base families, placed at their own spread, fall on the gamma curve, except Barabási-Albert, which lies above it: its oldest nodes are linked to each other, which raises $\rho$  beyond $\bar k(1 + cv^2)$. Trencséni~\cite{r20} calls this rise the degree distribution effect.

\subsubsection*{F.2 Spillover by distance}
\begin{figure*}[!tp]\centering
\includegraphics[width=0.88\textwidth]{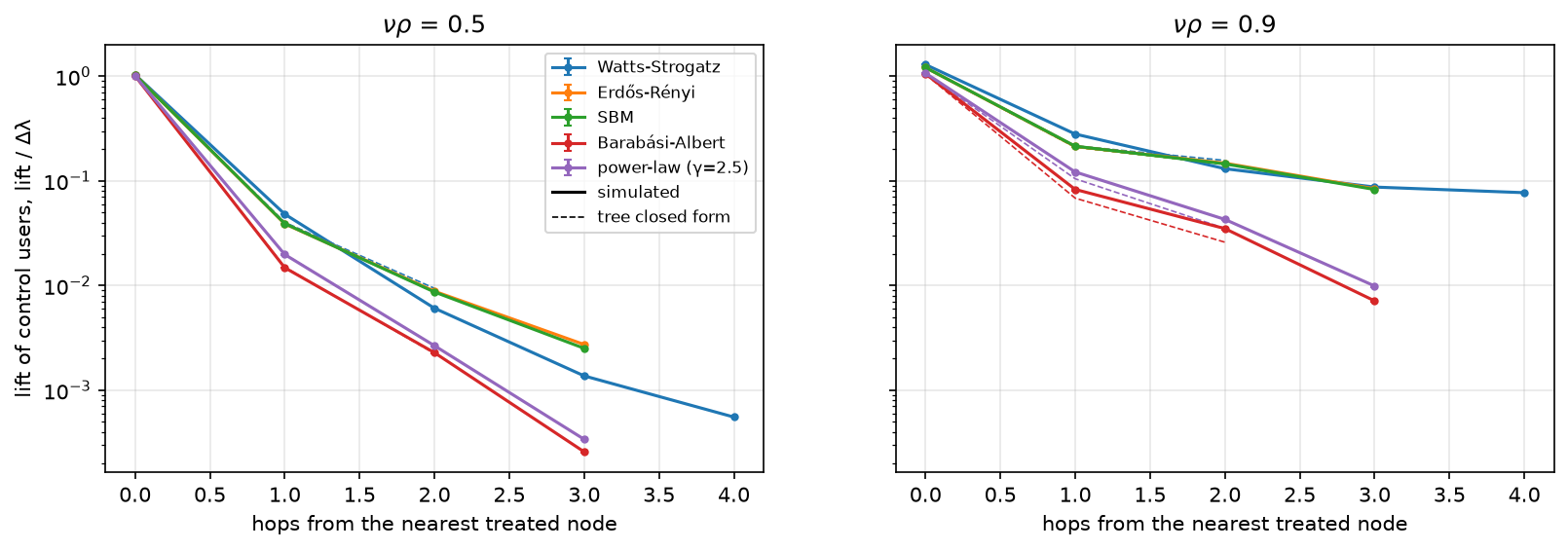}
\caption{Friends of treated users gain a little from the treatment, friends of friends much less. Lift of control users by hop distance from the nearest treated user, 2\% treated. Dashed: the tree closed form of Appendix B.2, which fails on Watts-Strogatz graphs, where neighbors share most of their neighbors.}\label{fig:fig5_3_spillover_hops}
\end{figure*}

Control users near treated users gain production; Trencséni~\cite{r20} calls this the spillover effect. The gain falls quickly with distance: friends of treated users gain a few percent of $\Delta\lambda$, friends of friends much less, and the decay per hop is slower at larger $\nu\rho$  (Figure \ref{fig:fig5_3_spillover_hops}). On a locally tree-like graph the response to one treated node falls by about $\nu$  per hop (B.2), because each node is reached by essentially one path. The lift at a given distance also counts the treated nodes farther away, so it falls more slowly, approaching $\nu\rho$  per hop at large distance.

\subsubsection*{F.3 Decomposition of the gap}

As shares of the GATE, the lift of the treated is $p + (1 - p)E[\mathrm{DM}]/\mathrm{GATE}$  and the spillover into control is $p(1 - E[\mathrm{DM}]/\mathrm{GATE})$. These are the two terms of the gap in Section 4.2, up to terms of order $1/N$, and the simulation follows both. In the decomposition of Hu, Li and Wager~\cite{r12} the average indirect effect is $\mathrm{AIE} = \mathrm{GATE} - \mathrm{ADE}$, and the missing indirect effect on the treated is about $(1 - p)\,\mathrm{AIE}$. In the terms of Trencséni~\cite{r20}, the attenuation ratio is the experiment dampening effect and the missing indirect effect is the intrinsic dampening effect.

\subsubsection*{F.4 Graph size}

Trencséni~\cite{r20} found that the effects do not change with graph size on Watts-Strogatz graphs. The same holds here up to $10^8$  nodes on the homogeneous families ($10^7$  for SBM), but not on the heavy-tailed ones (Figures \ref{fig:figM_2_size_fixed_nurho} and \ref{fig:figM_2b_size_fixed_nukbar}), whose largest degrees, and with them $\rho$, grow with $N$. On the power-law graph with the cutoff $k_{\max} = \sqrt N$, $\rho$  grows like $N^{1/4}$; a natural cutoff $N^{1/(\gamma - 1)}$  would make it grow faster. On Barabási-Albert graphs $\rho$  exceeds both terms of B.1, because the oldest nodes form a dense core. At a fixed $\nu\rho$  the heavy-tailed families therefore have a smaller $\nu$  on larger graphs, the attenuation ratio rises toward 1, and the Perron vector concentrates further on the hubs.

With $\bar k = 2$, one Erdős-Rényi and one Barabási-Albert graph per size up to $10^9$  nodes extend the size range of the multiplier and the attenuation ratio. On Barabási-Albert graphs $\rho$  follows $\sqrt{k_{\max}}$  and grows like $N^{1/4}$, and the attenuation ratio approaches 1. An Erdős-Rényi graph with $\bar k = 2$  is not homogeneous in the sense of this paper: its largest degrees set $\rho$, the Perron vector is concentrated on a few nodes, and the attenuation ratio rises slowly with $N$. The experiments of Sections 5 to 8 were not run at $10^9$  nodes.

\subsubsection*{F.5 Clustered assignment}

Trencséni~\cite{r20} calls the rise of the attenuation ratio under clustered assignment the clustering effect, and showed it with one contiguous treatment group on a Watts-Strogatz ring. Figure \ref{fig:figM_4_clusters_phi} shows it per family. The five families cover $\phi$  unevenly: METIS keeps few edges inside clusters on Erdős-Rényi and heavy-tailed graphs, many on SBM graphs when the clusters are as large as the blocks, and many on Watts-Strogatz graphs for all but the smallest clusters. SBM graphs with more edges between blocks fill the intermediate range of $\phi$, and there METIS clusters of the block size nearly match the true blocks.

\subsubsection*{F.6 Saturation and targeted treatment}
\begin{figure*}[!tp]\centering
\includegraphics[width=\textwidth]{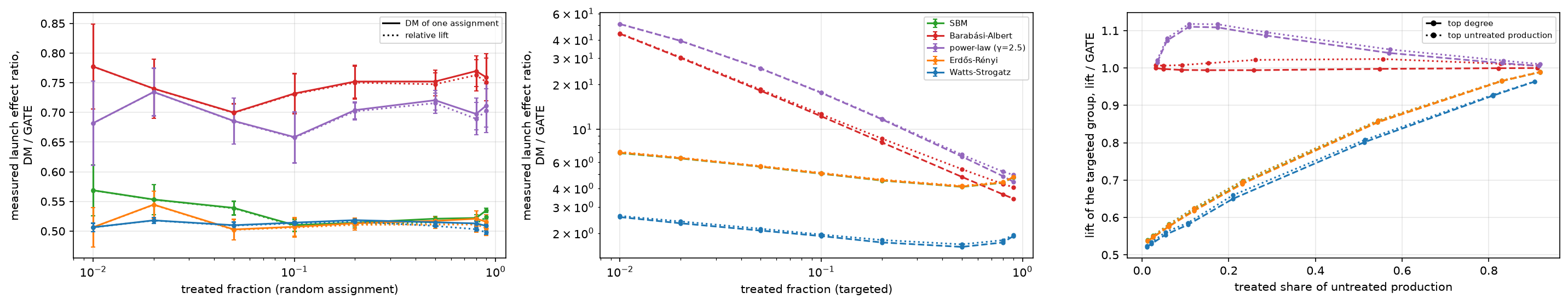}
\caption{With random assignment the treated fraction does not change the share of the launch effect the test reports (left). Treating the top users by degree (dashed) or by untreated production (dotted) mostly measures that they post more anyway (middle); their own lift over their untreated production is close to the GATE on the heavy-tailed families and between half of the GATE and the GATE on the homogeneous ones (right). $\nu\rho = 0.5$.}\label{fig:fig5_7_saturation}
\end{figure*}

Under random assignment the attenuation ratio does not depend on the treated fraction (Figure \ref{fig:fig5_7_saturation}, left), as the exact $E[\mathrm{DM}]$  of Section 2 requires; Trencséni~\cite{r20} calls the dependence on the treated fraction the experiment size effect. The points are single assignments, so they include the baseline difference between the arms, which is large on heavy-tailed graphs. Targeting the users with the most friends or the highest untreated production is not a randomized comparison (Figure \ref{fig:fig5_7_saturation}, middle). Most of the DM is then the difference in untreated production between the targeted users and the rest, which a pre-period comparison shows as well; relative to the GATE this difference grows like $\lambda/\Delta\lambda$. A randomized holdout of power users, or their own pre-period, measures the treatment; comparing them with everyone else measures their baseline advantage.

\subsubsection*{F.7 Build-up time}

When the Perron vector is flat, $\mathrm{GATE}(t)$  reaches 95\% of its limit after $\lceil\log 0.05/\log(\nu\rho)\rceil$  periods. The homogeneous families follow this exactly, and Barabási-Albert graphs are slightly faster. The DM settles in far fewer periods, except very close to the limit, where it also slows down, more on heavy-tailed graphs.

\subsubsection*{F.8 Cluster randomization by family}

Per family, at $\nu\rho = 0.3$, 0.6 and 0.9, the bias and the error of the DM and of the Hájek exposure estimator depend on cluster size as follows. On SBM graphs the best cluster size is the block size; on Watts-Strogatz graphs a wide range of sizes works; Erdős-Rényi and heavy-tailed graphs gain little. On heavy-tailed graphs the largest clusters raise the variance, and on Barabási-Albert graphs the mean DM itself changes with the cluster size, depending on which arm the cluster of hubs lands in. The Hájek estimator with one-hop exposure conditions (treated users with at least 75\% treated neighbors, control users with at most 25\%; Aronow and Samii~\cite{r1}) has a lower bias than the DM on the homogeneous families for clusters of up to 1,000, a higher one with very large clusters, and a much higher variance. Its one-hop exposure mapping is misspecified here, since outcomes depend on assignments at all distances.

\subsubsection*{F.9 Clusters from the wrong graph}
\begin{figure}[!htbp]\centering
\includegraphics[width=0.9\columnwidth]{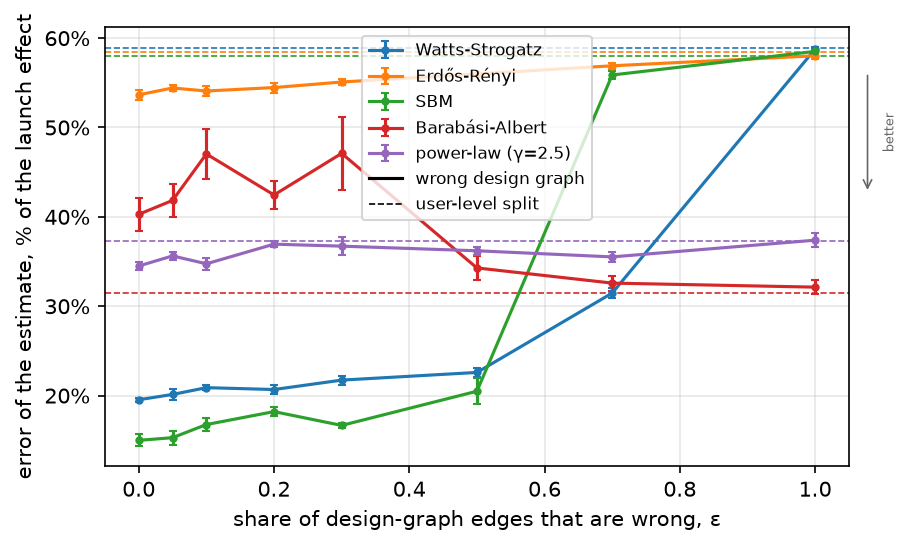}
\caption{Clusters built from an inaccurate friendship graph keep most of their benefit as long as the communities can still be found. Error relative to the GATE against the share of edges replaced at random.}\label{fig:fig8_2_wrong_graph}
\end{figure}

Clusters are usually computed from an approximation of the graph through which influence flows. Here clusters of 1,000 are computed on a copy of the graph in which a fraction $\varepsilon$  of edges is replaced by uniformly random edges, at $\nu\rho = 0.6$  (Figure \ref{fig:fig8_2_wrong_graph}). On Watts-Strogatz graphs the error rises slowly with $\varepsilon$. On SBM graphs it stays low until METIS stops recovering the blocks, then jumps to the level of a user-level split; B.4 with $\phi$  measured on the true graph reproduces the jump. On Erdős-Rényi and power-law graphs clusters help little at any $\varepsilon$. On Barabási-Albert graphs clusters from the true graph are worse than random ones: the variance added by clusters of hubs exceeds the bias removed.

\subsubsection*{F.10 Rollout estimators by family}

Per family and condition, the estimators of Section 6.2 compare as follows, together with a neighborhood regression on the one-hop and two-hop treated fractions. With the same graph and baseline production in both windows, the two-fraction slope and the historical baseline are close to the GATE, the plug-in less so, and the DM far off. A common drift in baseline production between the windows spoils both slopes: for a drift with standard deviation $\sigma_d$  the error is about $\sigma_d/((p_2 - p_1)\Delta\lambda/\lambda)$  of the GATE, so it grows as the effect or the difference in treated fractions shrinks. Wrong edges in the analysis graph spoil the plug-in, whose estimate of $\nu$  then moves in a different direction on each family. The neighborhood regression is far off in every condition: under a 50\% Bernoulli design the share of treated friends varies little between users, so the contrast between no friends and all friends treated extrapolates far outside the data. A randomized-saturation design would supply that variation and was not run.

\subsubsection*{F.11 Variance reduction}

CUPED with the pre-period outcome (Deng et al.~\cite{r7}) reduces the standard deviation of the DM under both designs. Its mean is the DM's, so it does not reduce the interference bias. Adding the treated share of a user's friends as a covariate changes what the treatment coefficient means: it becomes close to the direct effect, whatever the design. The all-treated contrast, the sum of the treatment and exposure coefficients, moves toward the GATE, at a much higher standard deviation.

\subsubsection*{F.12 Power near the stability limit}

Section 6.3 uses 100 assignments on each of 5 graphs, with $\nu\rho$  from 0.5 to 0.999, in runs with and without the effect (Figures \ref{fig:figM_6_critical_effects} and \ref{fig:figM_6b_critical_power}). The GATE grows by orders of magnitude while the treatment-induced DM stays near $\Delta\lambda$, so relative to mean production the lift shrinks toward zero. The detectable effect grows like $1/(1 - \nu\rho)$  when the baseline-imbalance term $cv_b^2$  of B.5 dominates the Poisson term $1/(W\lambda m)$, which it does near the limit on all five families; on a regular graph with homogeneous baseline production that term vanishes, and the detectable effect grows only like $(1 - \nu\rho)^{-1/2}$. With no effect the test rejects close to the nominal rate, with or without CUPED. CUPED keeps much of the power when each user's baseline is unchanged between the two windows, and little at the very limit.

\subsubsection*{F.13 Inference}

The reported posterior probability of a positive effect is the plug-in Gaussian approximation; it equals one minus the one-sided p-value for the same unit of analysis by construction. With no effect and cluster designs, analyses that treat users as independent reject far more often than the nominal 5\%, while cluster-level analyses are calibrated on every family. A Gamma-Poisson posterior that treats users as exchangeable also rejects too often, because it ignores differences between users. Only calibration under no effect was established; coverage when an effect is present was not. A calibrated interval still estimates the expected DM, not the GATE: even with the true standard error, a nominal 95\% interval must cover the GATE less often than 95\% of the time, and less often on a larger graph, because the standard error shrinks like $N^{-1/2}$  while the bias does not.

\subsubsection*{F.14 New edges}

Each treated user gets the boost $\Delta\lambda$  and forms Poisson($r$) new undirected edges; a second variant has only the edge channel (Figure \ref{fig:figM_7_new_edges} shows the first). New friends are friends of friends or uniformly random users, at $\nu\rho = 0.5$  on the original graph. The GATE grows quickly with $r$, much more on homogeneous than on heavy-tailed graphs, whose $\nu$  is smaller at the same $\nu\rho$. Under a 50/50 split half of a treated user's new edges go to control users, so the densification enters both arm means and largely cancels in the DM, which keeps only the difference in new edges between the arms. This is the missing indirect effect of Section 4.2, applied to the larger densification effect. Under the user-level split the attenuation ratio falls as $r$  grows, with or without the intrinsic boost. A cluster split reports more than the user-level split in both variants, but keeps most of the effect only when the new edges stay inside treated clusters, as with friends of friends on SBM and Watts-Strogatz graphs; on Erdős-Rényi and heavy-tailed graphs METIS finds no communities, and the cluster split is close to the user-level split.

\subsubsection*{F.15 New users}

Each treated user gets the boost $\Delta\lambda$  and invites Poisson($r$) new users; a new user connects to the inviter and to $q$  draws from the inviter's friends and produces $a\lambda$  (Figure \ref{fig:figM_7b_new_users} shows the estimators). The new users are added at once and the steady state is computed on the grown graph, so $r$  is the number of new users per treated user who have joined by the time the test is read. The total effect per original user is positive and grows with $r$. The per-user effect, averaged over all users including the new ones, is negative exactly when the total effect per original user is below $(N_{\mathrm{new}}/N)\bar c_0$, with $N_{\mathrm{new}}$  the number of new users; it can change sign as $r$  varies, so the figures show the total. The estimator of Section 7.2 reports more of the GATE as $r$  grows, because the new users' own production, which appears only in the treated arm, makes up more of it. Its shortfall is spillover to control users, larger when new users also befriend the inviter's friends ($q = 2$). Leaving the new users out reports little more than the $\Delta\lambda$  boost.

\subsubsection*{F.16 Threshold contagion}

The recursion of Gleeson and Cahalane~\cite{r10} (B.7) predicts a critical treated fraction $p^*$  that grows with the threshold $\theta$, nearly the same on Erdős-Rényi, Barabási-Albert and power-law graphs (Figure \ref{fig:figM_9_threshold}). On these graphs the transition is narrow and at $p^*$. On SBM graphs at high thresholds every run cascades just below $p^*$: the treated share varies between blocks, and some blocks start above $p^*$. On Watts-Strogatz graphs the transition is wider and earlier, more so at low thresholds: a node's neighbors are neighbors of each other, so a few seeds in one neighborhood start a front that the tree recursion does not describe. Without a cascade the DM is close to the GATE; with one, it is close to 0, and the average over assignments mixes the two.

\subsubsection*{F.17 Concave response}

Trencséni~\cite[Section 6]{r20} capped the peer term with a sigmoid and observed that production then no longer diverges; here the response is concave and the steady state is characterized. Figures \ref{fig:figM_8_concave_multiplier} and \ref{fig:figM_8b_concave_attenuation} show part of the grid. The local amplification $\rho(J)$  first rises with $\nu$  and then falls as saturation deepens, except for small $K$  on heavy-tailed graphs, where it is still rising at the end of the grid. The treated fraction makes no visible difference. A treatment ten times larger has a launch effect slightly below ten times that of the small one, as concavity requires. Cluster randomization helps as in Section 6.1, close to $(1 - \rho(J))/(1 - \rho(J)\phi)$, and little on graphs without communities. Mean production is no longer exactly linear in the treated fraction but nearly so: the two-fraction slope stays close to the GATE, and the historical baseline overshoots slightly for the large treatment, the direction that concavity predicts.

\subsection*{G. Related work}

Hudgens and Halloran~\cite{r13} defined direct and indirect effects under interference, and Aronow and Samii~\cite{r1} exposure mappings and their estimators; Hu, Li and Wager~\cite{r12} defined the average indirect effect, which with the direct effect sums to the total effect in a linear model. Ugander, Karrer, Backstrom and Kleinberg~\cite{r21} proposed graph cluster randomization, and Eckles, Karrer and Ugander~\cite{r8} showed that it reduces the bias of the difference in means under linear models. Estimators of the total effect use a historical baseline (Yu, Airoldi, Borgs and Chayes~\cite{r25}) or a staggered rollout (Cortez, Eichhorn and Yu~\cite{r6}); Deng, Xu, Kohavi and Walker~\cite{r7} introduced CUPED. The model used here, in which a user responds to the sum of their neighbors' content, is estimated by spatial two-stage least squares (Kelejian and Prucha~\cite{r16}; Bramoullé, Djebbari and Fortin~\cite{r4}).

\FloatBarrier
\makeatletter\global\@colroom\@colht\global\vsize\@colht\makeatother

\end{document}